\documentclass[twocolumn]{jpsj3} 
\usepackage{bm} 
\usepackage{graphicx,color}%
\usepackage{amsmath,amsfonts,amsthm,amssymb,slashbox} 
\newcommand{\br}{\mbox{\boldmath $r$}} 
\newcommand{\bR}{\mbox{\boldmath $R$}}

\title{
Comparison of {\em Ab initio} Low-Energy Models for LaFePO, LaFeAsO, BaFe$_2$As$_2$, LiFeAs, FeSe and FeTe:  
 Electron Correlation and Covalency 
} 
 
\author{Takashi \textsc{Miyake}$^{1,3,4}$
\thanks{Electronic mail: t-miyake@aist.go.jp}, Kazuma \textsc{Nakamura}$^{2,3,4}$, Ryotaro \textsc{Arita}$^{2,3,4}$,  \\ and Masatoshi \textsc{Imada}$^{2,3,4}$}  
 
\inst{ 
$^{1}$ Research Institute for Computational Sciences, AIST, Tsukuba 305-8568, Japan \\ 
$^{2}$ Department of Applied Physics,  
 University of Tokyo, 7-3-1 Hongo, Bunkyo-ku, Tokyo 113-8656, Japan \\  
$^{3}$ Japan Science and Technology Agency, CREST, Honcho, Kawaguchi, Saitama 332-0012, Japan \\ 
$^{4}$ Japan Science and Technology Agency, TRIP, Honcho, Kawaguchi, Saitama 332-0012, Japan 
} 
 
\recdate{\today} 
 
\abst{ 
{\bf Effective low-energy Hamiltonians for several different families of iron-based superconductors are compared after deriving them from the downfolding scheme based on first-principles  
calculations. 
Systematic dependences of the derived model parameters on the families are elucidated, many of which are understood from the systematic variation of the covalency between Fe-3$d$ and pnictogen-/chalcogen-$p$ orbitals.}
First, 
LaFePO, LaFeAsO (1111), BaFe$_2$As$_2$ (122), LiFeAs (111), FeSe and FeTe (11) have overall similar band structures 
near the Fermi level, where the total widths of ten-fold Fe-3$d$ bands are mostly around 4.5 eV.  
However, the derived effective models of the ten-fold iron-3$d$ bands ($d$ model) for FeSe and FeTe have substantially larger effective onsite Coulomb interactions $U \sim 4.2$ and $\sim 3.4$ eV, respectively, after the screening by electrons on other bands and after averaging over orbitals, 
as compared to $\sim$ 2.5 eV for LaFeAsO. 
The difference is similar in the effective models containing $p$ orbitals of As, 
Se or Te ($dp$ or $dpp$ model), where $U$ ranges from $\sim$ 4 eV for the 1111 family to 
$\sim 7$ eV for the 11 family. The exchange interaction $J$ has a similar tendency.
The family dependence of models indicates a wide variation ranging from 
weak correlation regime (LaFePO) to substantially strong correlation regime (FeSe). 
The origin of the larger effective interaction 
in the 11 family is ascribed to smaller spread of the Wannier orbitals generating larger bare interaction, and to fewer screening channels
by the other bands.  
This variation is primarily derived from 
the distance $h$ between the pnictogen/chalcogen position and the Fe layer: The longer $h$ for the 11 family generates more ionic character of the 
 bonding between iron and anion atoms, 
 while the shorter $h$ for the 1111 family leads to 
 more covalent-bonding character, the larger spread of the Wannier orbitals, and more efficient screening by the anion $p$ orbitals. 
The screened interaction of the $d$ model is strongly orbital dependent, which is also understood from the Wannier spread. 
The $dp$ and $dpp$ models show much weaker orbital dependence.
The larger $h$ also explains why the ten-fold 3$d$ bands for the 11 family are more entangled with the smearing of the ``pseudogap" structure above the Fermi level seen in the 1111 family. 
While the family-dependent semimetallic splitting of the bands primarily consists of $d_{yz}/d_{zx}$ and $d_{x^2-y^2}$ orbitals, 
the size of the pseudogap structure is controlled by the hybridization between these orbitals and $d_{xy}/d_{3z^2-r^2}$: 
A large hybridization in the 1111 family generates a large ``band-insulating"-like pseudogap 
({\it hybridization gap}), whereas a large $h$ in the 11 family weakens them,
resulting in a ``half-filled" like bands of orbitals.
This may enhance strong correlation effects in analogy with Mott physics and causes the orbital selective crossover in the three orbitals. 
On the other hand, the 
geometrical frustration $t'/t$, inferred from the ratio of the next-nearest transfer $t'$ to the nearest one $t$ of the $d$ model is
relatively larger for the 1111 family than the 11 one. 
The models comprehensively derived here may serve as a firm starting basis of understanding both common and diverse properties of the iron-based superconductors including magnetism and superconductivity.} 
\kword{%
first-principles calculation, effective Hamiltonian, downfolding, constrained RPA method, LaFePO, LaFeAsO, BaFe$_2$As$_2$, LiFeAs, FeSe, FeTe, oxypnictide, oxychalcogenide, high-temperature superconductivity} 
 
\begin{document} 
\maketitle 
  
\section{Introduction} 
Recently a new class of superconducting compounds including iron element 
has been discovered.\cite{Hosono}  In all the cases of this class, Fe-3$d$ conduction electrons  
are likely to form Cooper pairs and responsible for the superconductivity. However,  
the mechanism of superconductivity is not well understood and is under extensive debates.  
In the family with ZrCuSiAs-type structure (called 1111 hereafter),  
SmFeAs(O,F) has shown the highest superconducting critical temperature $T_{\rm c} \sim 56$ K (ref.~\citen{Ren})  
when fluorine is substituted for $\sim 20\%$ of oxygen as electron doping, while BaFe$_2$As$_2$ with ThCr$_2$Si$_2$-type structure (called 122)  
has indicated the highest $T_{\rm c}\sim 38$ K, when potassium is substituted for $\sim 40 \%$ of Ba as hole doping.\cite{Rotter}  There exist another simpler compounds  LiFeAs and NaFeAs (called 111) reported as the PbFCl-type tetragonal structure, indicating $T_c \sim 18$ K.\cite{LiFeAs_Wang,LiFeAs_Pitcher,LiFeAs_Tapp} 
Another family of binary compounds FeSe$_x$Te$_{1-x}$ (called 11) also shows 
superconductivity at $T_{\rm c}$ higher than 10 K (refs.~\citen{11,Li}) and has reached 37 K under pressure (7 GPa).\cite{Margadonna}  
 
It is highly desired to understand what are common and what are family dependent, 
from detailed electronic structure of these 
four families, for the purpose of establishing the basis for revealing mechanisms of superconductivity and magnetism. 
In this report, we present effective low-energy models of these families derived from first principles and compare them. 
 
A common known feature of iron-based superconductors  
is that the antiferromagnetic order appears close to the superconducting region except for the 111 family.  
The mother compound of the 1111-type, for example, LaFeAsO shows antiferromagnetic  
long-range order of the stripe type below $T_N\sim 130$ K with the Bragg point at $(\pi,0)$  
in the extended Brillouin zone 
(corresponding to $(\pi,\pi)$ in the reduced Brillouin zone).\cite{Mook} The antiferromagnetic ordered  
moment $\sim$0.36 $\mu_B$ as compared to the nominal saturation  
moment 4 $\mu_B$ for the high-spin 3$d^{\rm 6}$ state is unexpectedly  
small, implying large quantum fluctuations arising from  
electron-correlation effects or dominating itinerancy with subtlety of competing ground states. 
On the other hand, the 122-type (BaFe$_2$As$_2$) shows a relatively large ordered moment 
 $\sim 1.1$ $\mu_B$ (refs.~\citen{Huang}~and~\citen{Kofu}) and the 11-type (FeTe) indicates even larger ordered moment  
$\sim 2.25$ $\mu_B$ at a different Bragg point, $(\pi/2,\pi/2)$ 
in the extended Brillouin zone.\cite{Li}  
Even for the 1111 family, NdFeAsO shows larger ordered moment ($\sim 0.9$ $\mu_B$),  
although the moment is apparently reinforced by 
Nd moment ($\sim 1.55$ $\mu_B$).\cite{Nd}  
 
Conventional density-functional calculations with the local density  
approximation (LDA) or the generalized gradient approximation (GGA) 
have clarified that bands originating from ten-fold  
degenerate iron-3$d$ orbitals in a unit cell containing two Fe atoms are close to the Fermi  
level.   
The LDA calculations of the 1111-type,~\cite{Lebegue,Singh,Hirschfeld,Terakura,Ma,Kuroki,Nekrasov1111} 122-type,~\cite{Singh_122,Nekrasov122} 
111-type,~\cite{Nekrasov111} and 11-type compounds~\cite{Subedi_SinghFeSe,MaFeSe} 
show a very similar band structure for all of the above families, 
where small electron pockets around M point and hole pockets around $\Gamma$ point  
lead to a semimetallic Fermi surfaces.  The local spin density approximation (LSDA) also commonly predicted the  
antiferromagnetic order for mother materials.\cite{Hirschfeld,Terakura,Ma}  
The stripe-type antiferromagnetic order is  
correctly reproduced for the 1111-type.\cite{Terakura,Ma} However, the calculated ordered  
moment obtained so far is large and ranges between 1.2 and 2.6 $\mu_B$,\cite{Hirschfeld,Terakura,Ma,Mazin2}   
in contrast to much smaller ordered moment discussed above. 
It is unusual to observe the ordered moment smaller than the LSDA result. 
On the other hand, the bicolinear order for FeTe is reproduced  
in the LSDA with more or less consistent ordered moment ($\sim 2.25$ $\mu_{\rm B}$) with the experimental  
results.\cite{MaFeSe} 
Broad peak structures of magnetic Lindhard function calculated  
by using the LDA/GGA Fermi surface suggest competitions of several  
different ordering tendencies.\cite{Mazin,Dong,Kuroki,KurokiPRB,Yildirim}

The role of electron correlation based on the realistic grounds is under a strong debate.\cite{Nakamura,Haule,Anisimov,Craco} 
Relatively small fraction of the Drude weight~\cite{Boris,Hu,Yang_Timusk,Basov} together with bad metallic (or semiconducting)  
behavior~\cite{Hosono,Chen}  
indicate substantial electron correlation effects. 
Antiferromagnetic orders and fluctuations themselves revealed by the nuclear magnetic resonance and other probes 
near the superconducting phases also indicate,  
in any case, some electron correlation effects play a role.~\cite{Hosono-Nakai-Ishida,NMR}  
Diversity of the ordered moment ranging from 0.36 $\mu_{\rm B}$ to 2.26 $\mu_{\rm B}$ is remarkable in terms of the  
similar band structure with semimetallic small pockets of the Fermi surface. Ordering vector of the antiferromagnetic order introduced above  
depending on the compounds further suggests that the correlation effects and its subtlety  
are beyond the simple nesting and weak coupling picture. In fact, recent fluctuation exchange calculation suggests that the self-energy effect with subtle multiband structure near the Fermi level cast a serious suspicion on the validity of the nesting picture.\cite{AritaIkeda} 

Angle resolved photoemission spectroscopy~\cite{Lu_Shen_1111,Yi_Shen_122} 
 has shown some correspondence to the LDA result of Singh {\it et al}.\cite{Singh_122} 
Fe-2$p$ core-level spectra of X-ray photoemission suggest rather itinerant character.\cite{Malaeb,Shen_Xray} 
On the other hand, some role of moderate electron correlations has also been claimed.\cite{Takahashi,AnisimovPhysica} 
For FeSe, soft-Xray photoemission results~\cite{Shin,Yamasaki} appears to show a deviation from the LDA results suggesting a splitting of the coherent band near the Fermi level from the incoherent part arising from the correlation effect as we discuss in Sec.\ref{Summary}.  

In the superconducting phase, even the pairing symmetry itself is highly controversial and no consensus has been reached. 
Although nodeless superconductivity is suggested~\cite{Kaminski,Ding,Hashimoto}, temperature dependence of nuclear-magnetic-relaxation  
time $T_1$ below $T_{\rm c}$ roughly scaled by $T^{-3}$ without the Hebel-Slichter peak
implies unconventional superconductivity driven by nontrivial electron-correlation  
effects.~\cite{NMR}  For example, orbital dependent gaps with sign-changing and fully-gapped $s\pm$ symmetry has been proposed.~\cite{Hosono-Nakai-Ishida}
The gradual suppression of the superconducting transition temperature by Co doping into the Fe site was reported to be explained by
strong antiferromagnetic fluctuations near the metal-insulator boundary producing an effect on the $s$-wave singlet pairing without the sign change.\cite{Sato1,Sato2,Onari}   
Although overall experimental results suggest noticeable correlation  
effects, realistic roles of electron correlations on theoretical  
grounds are not well established and controversial.

The effective electron Coulomb repulsion of the 1111 family for the models of the ten bands of Fe-3$d$ orbitals 
has been estimated by Nakamura {\it et al.} from first principles 
by applying the downfolding  
scheme to eliminate other band degrees of freedom~\cite{Nakamura} along the line of the three-stage scheme.~\cite{aryasetiawan04,Imai1,Imai2}  
The ratio of the Hubbard onsite interaction $U\sim 3$ eV to the nearest neighbor transfer $t\sim 0.3$ eV in the downfolded model 
is estimated to be $U/t\sim 10$ 
with the fivefold orbital degeneracy, 
indicating a moderately strong correlation. 
This identification has also been supported for the case of the 122-type.\cite{Anisimov_Vollhardt}
Here it should be noted that the effective Coulomb interaction estimated in these works as well as 
in the present paper for the low-energy model is not directly the same as the interaction derived in experimental probes such as the X-ray photoemission.\cite{Malaeb,Shen_Xray}  
This is because, in the model parameters here, the screening arising from the polarization 
within the low-energy degrees of freedom (Fe-$3d$ bands in ref.~\citen{Nakamura}) is excluded 
as we describe in the next section, because this screening effect should be considered 
when the low-energy model is solved. 
On the other hand, in the experiment, the interaction effect appears 
as a whole consequence after the full screening. We will further discuss this issue in Sec.\ref{Summary}.

When we consider the puzzling diversity of magnetic and transport properties among the 1111, 122, 111 and 11-types 
in spite of the apparent similarity of the band structure by the LDA, it is crucially important to elucidate 
the origin of the difference and diversity from a unified first-principles calculations by taking into account electron correlations properly.  
In this paper, we extend the work for the 1111 
family~\cite{Nakamura,miyake08b,aichhorn09}
and derive effective low-energy models of the 1111 as well as 122, 111 and 11 families on a unified grounds.  
We further classify and compare effective models toward the comprehensive understanding of the electron correlation effects.  
This comparison on the diversity is also important for the understanding of the superconducting mechanism. 
A key quantity is the pnictogen/chalcogen height, $h$. 
It was pointed out in the early stage that the electronic band structure is altered significantly by changing $h$ \cite{Singh,vildosola08}. 
There is also a previous work claiming that the spin and the charge susceptibility are sensitive to $h$ \cite{KurokiPRB}. 
The present study reveals that the strength of electron correlation is determined by the spatial extent of the 
Wannier orbitals and the strength of screening effect, and both of them are affected by $h$.
 
In Sec.2 we describe our method. Sec. 3 describes the derived effective models for LaFePO, LaFeAsO, BaFe$_2$As$_2$, LiFeAs, FeSe and FeTe.  
We present effective models both for the ten-band model for the iron-3$d$ Wannier orbitals ($d$ model) and the model including $p$ orbitals of 
P, As, Se or Te ($dp$ 
or $dpp$ model). Sec. 4 is devoted to 
summary and discussions.  
 
\section{Method} 
We derive the low-energy models  
by a combined constrained 
random-phase-approximation (cRPA) (ref.~\citen{aryasetiawan04})  
and maximally localized Wannier function 
(MLWF) (refs.~\citen{marzari97}~and~\citen{souza01}) method.  
This combination has been recently developed and successfully applied to  
the 3$d$ transition metals,\cite{miyake08,Miyake} 
their compounds~\cite{miyake08}  
including LaFeAsO,\cite{Nakamura,miyake08b,aichhorn09}   
organic conductors,\cite{nakamura09} and zeolites.\cite{nakamura09-2} 
 
The first step of the method is a standard band structure calculation  
in the framework of density functional 
theory (DFT).~\cite{hohenberg64,kohn65} 
We then choose target bands around the Fermi level 
and 
extract them following the MLWF procedure,  
which defines the one-body part of the low-energy model.  
The transfer integral is obtained by taking the matrix element of  
the Kohn-Sham Hamiltonian, ${\cal{H_{{\rm KS}}}}$, in the MLWF basis,  
\begin{equation} 
t_{mn}({\bR}) = \langle \phi_{m{\mathbf 0}} | {\cal{H_{{\rm KS}}}} | \phi_{n{\bR}} \rangle 
\;, 
\end{equation} 
where $\phi_{n{\bR}}({\br})$ is the MLWF  
centered at the site ${\bR}$ for the $n$-th orbital.  
 
To evaluate the effective interaction parameters,  
partially screened Coulomb interaction at zero frequency, 
$W_r(\br,\br^\prime; \omega=0)$,  
is calculated in the cRPA with the constraint that 
screening channels inside the target bands are cut out. 
This constraint is imposed to avoid double counting of 
the screening effects; 
they are considered later when the derived model is solved.  
The effective Coulomb interaction $U$ and 
exchange interaction $J$ are orbital dependent.  
Their matrix elements are given by  
\begin{eqnarray} 
U_{mn}({\bR}) &=& \langle \phi_{m{\mathbf 0}} \phi_{m{\mathbf 0}} | W_r |  
\phi_{n{\bR}} \phi_{n{\bR}} \rangle 
\; , \\ 
J_{mn}({\bR}) &=& \langle \phi_{m{\mathbf 0}} \phi_{n{\mathbf 0}} | W_r |  
\phi_{n{\bR}} \phi_{m{\bR}} \rangle 
\;, \\ 
\langle \phi_{i} \phi_{j} | W_r | \phi_{k} \phi_{l} \rangle &\equiv& 
\int\int  
\phi_{i}^*({\br}) \phi_{j}({\br}) 
W_r({\br},{\br^\prime}; \omega=0)  
\nonumber \\ &&\times 
\phi_{k}^*({\br^\prime}) \phi_{l}({\br^\prime}) 
d{\br} d{\br^\prime}  
\; . 
\end{eqnarray} 

One problem in the cRPA method is that, in the case of entangled band structure, 
it is not clear which screening process is to be excluded.  
This is indeed the case in all the materials studied in this work except for FeSe.  
Extension of the cRPA technique for entangled band structure has been  
proposed very recently.\cite{Miyake} 
In this method, the Hilbert space is divided into two parts;   
low-energy space spanned by the MLWFs and the rest of the space.  
Neglecting hybridization between the two spaces and   
diagonalizing the Kohn-Sham Hamiltonian in each space,  
disentangled band structure is obtained.  
Screening channel inside the target space is well-defined 
in the disentangled band structure,  
hence the cRPA calculation can be done without ambiguity.  
The present study serves as a good application of this technique.  

 The band structure calculation is based on the full-potential LMTO implementation.\cite{methfessel00}   
The exchange-correlation functional is the local density approximation  
of the Cepeley-Alder type~\cite{ceperley80} and spin-polarization is neglected.  
The cRPA calculation uses a mixed basis consisting of products of two atomic  
orbitals and interstitial plane waves.\cite{schilfgaarde06}  
The self-consistent LDA calculation is done for the $12 \times 12 \times 6$ $k$-mesh,  
and $20 \times 20 \times10$ $k$ points are sampled for the density of states.  
Both the partially screened Coulomb interaction  
and the MLWF setup use the 4$\times$4$\times$4 mesh.  
More technical details are found elsewhere.\cite{miyake08}  
 
The cRPA calculations were also performed with another {\em ab initio}  
band-structure code based on plane-wave basis set,  
{\em Tokyo Ab initio Program Package},\cite{Ref_TAPP}  
for critical comparisons with the FP-LMTO results.  
Density-functional calculations with LDA  
within the parameterization of Perdew-Wang~\cite{Ref_PW92} 
were performed with the  
Troullier-Martins norm-conserving pseudopotentials~\cite{Ref_PP1} 
in the Kleinman-Bylander representation.\cite{Ref_PP2} 
Iron pseudopotential was constructed under the reference configuration 
(3$d$)$^{7.0}$(4$s$)$^{0.8}$(4$p$)$^{0.0}$ by employing 
the cutoff radius for the 3$d$ state at 1.3 Bohr 
and for 4$s$ and 4$p$ states at 2.1 Bohr,  
with supplementation by the partial core correction of 
cutoff radii of 0.6 Bohr.   
The cutoff energies in wavefunctions and charge densities were set to 
 100 Ry and 900 Ry, respectively, and a 5$\times$5$\times$5 
 $k$-point sampling was employed.  
The polarization function was expanded in plane waves with 
 an energy cutoff of 20 Ry and the total number of bands 
 considered in the polarization calculation was set to 130.  
The Brillouin-zone integral on wavevector was evaluated by 
 the generalized tetrahedron method.\cite{Fujiwara}  
The additional terms in the long-wavelength polarization function 
due to nonlocal terms in the pseudopotentials were 
explicitly considered following ref.~\citen{Louie}.  
A problem due to the singularity in the Coulomb interaction, 
 in the evaluation of the Wannier matrix elements, 
 $U_{mn}({\bR})$ and $J_{mn}({\bR})$ in 
 Eqs.~(2) and (3), was treated in the manner 
 described in ref.~\citen{Louie}. 
We checked that these conditions give well converged results. 
 
A dependence of the resulting screened onsite parameters on 
cutoff radius of the iron pseudopotential was carefully checked 
through calculations with different choices 
of the cutoff radii of 1.3, 1.7, and 2.1 Bohr.  
We found that these different choices make no discernible difference 
in the resulting values and the difference is 
less than 0.1 eV at maximum; for example, for $U_{xy}$ of LaFeAsO,  
the value is 
3.14
eV for $r_c$ = 1.3 Bohr, 
3.20 eV for $r_c$ = 1.7 Bohr, and 
3.17 eV for $r_c$ = 2.1 Bohr.   
 Notice that differences in the values are not necessarily 
 monotonic.     

Now the polarization effects from the other bands far from the Fermi level are considered 
in the cRPA yielding the screened Coulomb interaction of the target bands.  
The degrees of freedom for the other bands are eliminated, leaving the low-energy degrees of freedom 
within the target bands only.  Thus obtained low-energy effective Hamiltonian has the form
\begin{eqnarray}
&&\mathcal{H}
= \sum_{\sigma} \sum_{ij} \sum_{nm}  
  t_{mn} ({\bR_i}-{\bR_j})
                   a_{in}^{\sigma \dagger} 
                   a_{jm}^{\sigma}   \nonumber \\
&&+ \frac{1}{2} \sum_{\sigma \rho} \sum_{ij} \sum_{nm} 
  \biggl\{ U_{mn}({\bR}_i-{\bR}_j) 
                   a_{in}^{\sigma \dagger} 
                   a_{jm}^{\rho \dagger}
                   a_{jm}^{\rho} 
                   a_{in}^{\sigma} \nonumber \\ 
&&+J_{mn}({\bR}_i-{\bR}_j) 
\bigl(\!a_{in}^{\sigma \dagger} 
      \!a_{jm}^{\rho \dagger}
      \!a_{in}^{\rho} 
      \!a_{jm}^{\sigma} 
  \!+\!a_{in}^{\sigma \dagger} 
     \!a_{in}^{\rho \dagger}
     \!a_{jm}^{\rho} 
     \!a_{jm}^{\sigma}\bigr)\! \biggr\}, 
\label{eq:H}                
\end{eqnarray}
where $a_{in}^{\sigma \dagger}$ ($a_{in}^{\sigma}$) 
is a creation (annihilation) operator of an electron with 
spin $\sigma$ in the $n$th MLWF centered on 
Fe atom at $\bR_i$. 
In Sec. \ref{Results} the target bands left in the low-energy bands are 
either Fe-3$d$ bands ($d$ model) or Fe-3$d$ as well as $p$ bands of pnictogen and oxygen (or chalcogen) ($dp/dpp$ model).

\section{Results}\label{Results}
\subsection{Band structure and density of states}
\begin{figure*}[htbp] 
\begin{center} 
\includegraphics[width=0.87\textwidth]{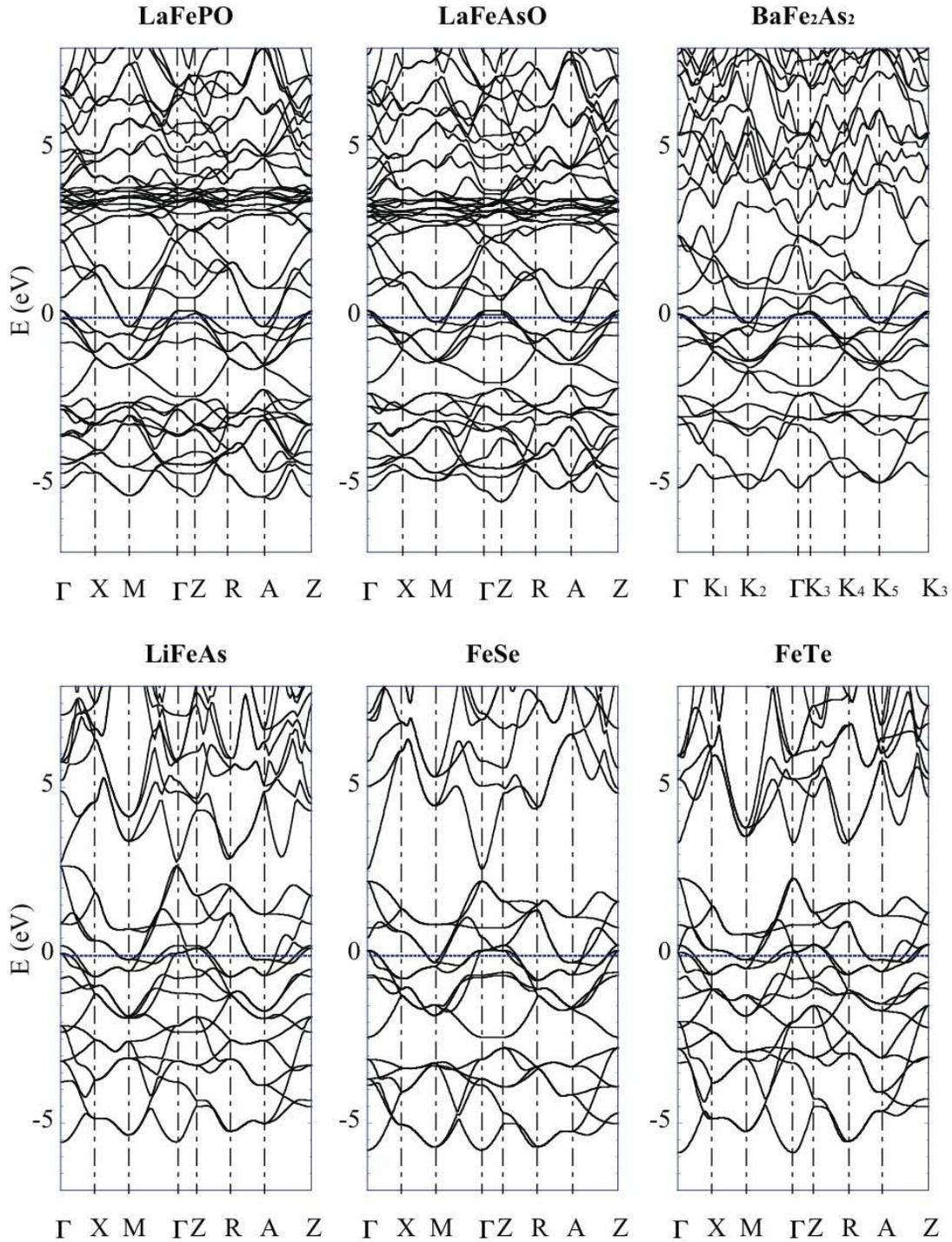}
\end{center} 
\caption{ 
Electronic band structures 
of six iron-based superconductors obtained by DFT-LDA. 
The $K_1-K_5$ points in BaFe$_2$As$_2$ are 
$
K_1 = \frac{2\pi}{a}(\frac{1}{2},0,0), 
K_2 = \frac{2\pi}{a}(\frac{1}{2},\frac{1}{2},0), 
K_3 = \frac{2\pi}{a}(0,0,\frac{a}{2c}),
K_4 = \frac{2\pi}{a}(\frac{1}{2},0,\frac{a}{2c}),
K_5 = \frac{2\pi}{a}(\frac{1}{2},\frac{1}{2},\frac{a}{2c}), 
$
respectively.
Energy is measured from the Fermi level. 
} 
\label{fig:band} 
\end{figure*} 
\begin{figure}[t] 
\begin{center} 
\includegraphics[width=0.46\textwidth]{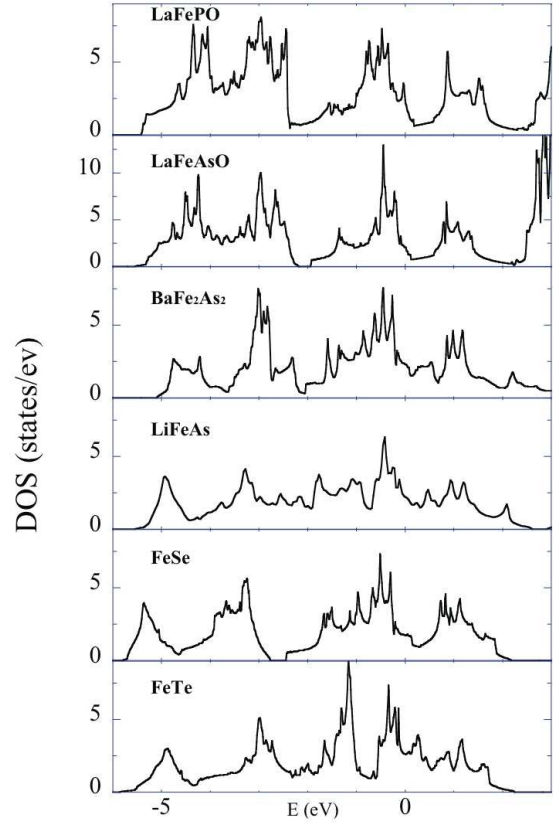}
\end{center} 
\caption{ 
Density of states 
of six different iron-based superconductors obtained by DFT-LDA. 
Number of states is counted for one-half formula unit in BaFe$_2$As$_2$, 
and for one formula unit in the other materials. 
Energy is measured from the Fermi level. 
} 
\label{fig:dos} 
\end{figure}
\begin{figure}[htbp] 
\begin{center} 
\includegraphics[width=0.45\textwidth]{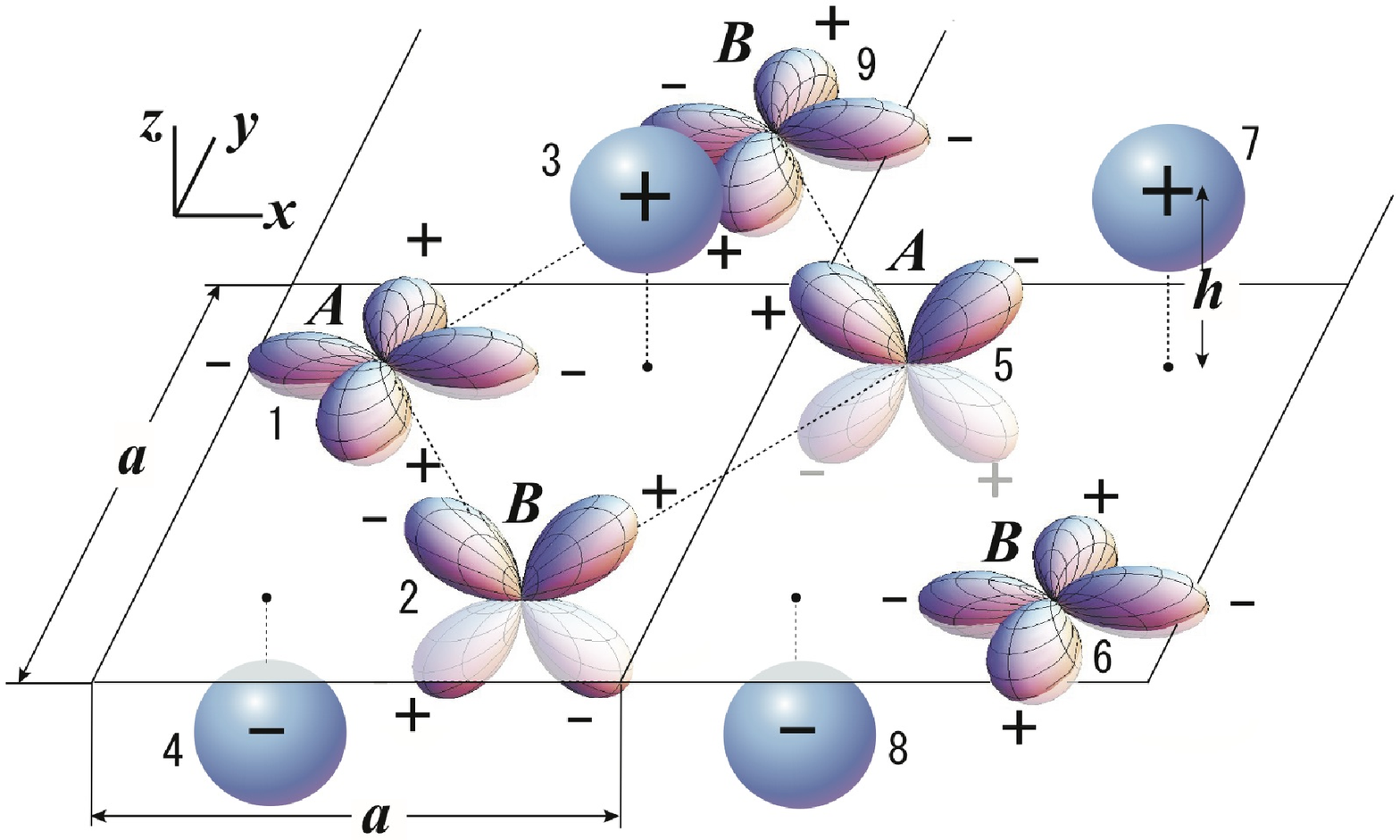}
\end{center} 
\caption{Color:
Schematic view of the structure for iron-based superconductors.
Positions of pnictogen (As or P) or chalcogen (Se or Te) atoms are illustrated by spheres at the sites 3, 4, 7 and 8 located
either above the iron layer (semitransparent sheet) depicted by the
plus sign inside the spheres or below the iron layer shown by the minus sign.
Iron sites are illustrated by examples of $x^2-y^2$ (the sites 1, 6 and 9) or $zx$ (the sites 2 and 5) symmetries of the iron 3$d$ orbitals,
where $A$ and $B$ indicate the sublattice indices in the unit cell.
The signs beside the $d$ orbitals give the rule of the local gauge employed in the present paper, where $xy, x^2-y^2$ and $3z^2-r^2$ orbitals have the uniform phases, whereas $zx$ and $yz$ orbitals change the sign
alternatingly depending on the $A$ or $B$ sublattices as is seen in the sites 2 and 5. 
With this local gauge, for example, through the hybridizations 
with the pnictogen-/chalcogen-$p$ orbitals, the transfers (including their signs) between the sites 1 and 2 become the same as those between 9 and 5 
in the $d$ model, 
which exemplifies the invariance with the translations $(\pm a/2, \pm a/2)$
in the $ab$ plane,
so that the model can have this translational symmetry apparently higher
than the real symmetry given by the primitive translation $(a, a)$ in the $ab$ plane.  
}
\label{fig:geometry} 
\end{figure}

 
Figures \ref{fig:band} and \ref{fig:dos} show the band structures  
and densities of states, 
respectively, of  
LaFePO, LaFeAsO, BaFe$_2$As$_2$, LiFeAs, FeSe and FeTe  
in the experimental geometry (Table~\ref{tab:geometry}).  
The overall feature of the band structure is common in all the compounds,
reflecting the common existence of the Fe layer sandwiched by the pnictogen/chalcogen atoms as is illustrated in Fig.~\ref{fig:geometry}.  
In Fig.~\ref{fig:band} we see entangled ten 
bands having strong Fe-3$d$ character  
located near the Fermi level.\cite{Lebegue,Singh,Hirschfeld,Terakura,Ma,Kuroki,Singh_122,Subedi_SinghFeSe,MaFeSe,Nekrasov1111,Nekrasov122,Nekrasov111}  
The bandwidth ranges from 4.4-4.6 eV (LaFeAsO, BaFe$_2$As$_2$, FeSe, FeTe) to 
4.9-5.1 eV (LiFeAs and LaFePO). 
The systems are metallic in the LDA with  
electron pockets around the M-A line  
and hole pockets formed around the $\Gamma$-Z line.  
Below the $d$ band are three states, 
which are mainly of P-/As-/Se-/Te-$p$ character.  
\begin{table}[h]
\caption{
Lattice parameters used in the present work.
Here, $h$ is the distance between the pnictogen/chalcogen atom and the Fe plane as is illustrated in
Fig.~\ref{fig:geometry}.}
\ 
\label{tab:geometry}
\begin{tabular}{lrrrl}
\hline 
\hline
& $a$ ($\AA$) & $c$ ($\AA$) & $h$ ($\AA$)  \\
\hline
LaFePO & 3.9636 & 8.5122 & 1.1398 & ref.~\citen{LaFePO} \\
LaFeAsO & 4.0353 & 8.7409 &  1.3216 & ref.~\citen{Hosono} \\
BaFe$_2$As$_2$ & 3.9625 & 13.0168 & 1.3602 & ref.~\citen{Rotter} \\
LiFeAs & 3.7764 & 6.3568 & 1.5075 & ref.~\citen{LiFeAs_Pitcher} \\ 
FeSe & 3.7738 & 5.5248 & 1.4652 & ref.~\citen{margadonna08} \\
FeTe & 3.8123 & 6.2517 & 1.7686 & ref.~\citen{Li} \\
\hline
\hline
\end{tabular}
\end{table}

\begin{figure*}[htbp] 
\begin{center} 
\includegraphics[width=0.44\textwidth]{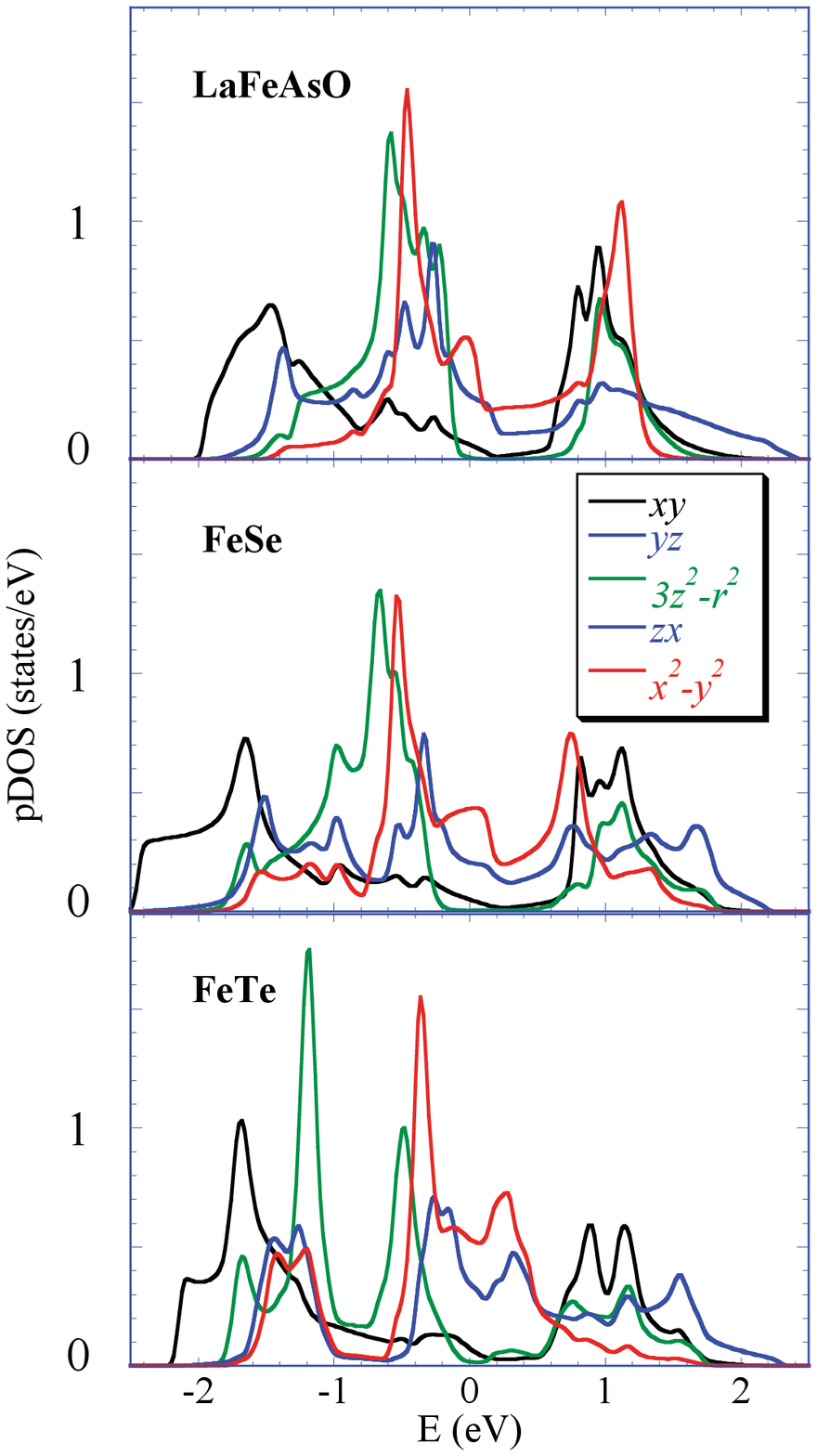}
\includegraphics[width=0.44\textwidth]{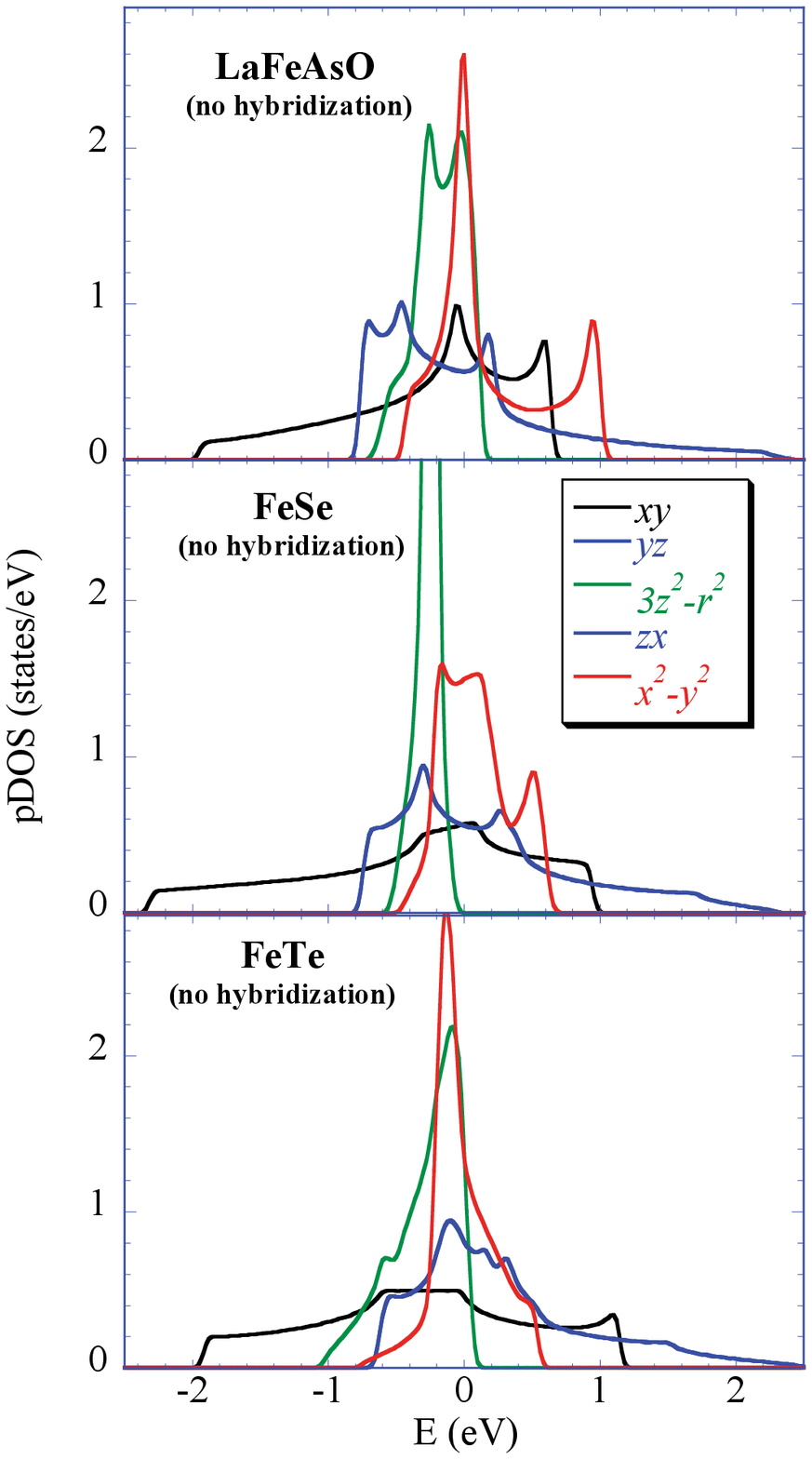}
\end{center} 
\caption{Color:
(Left panel) Partial density of states 
of LaFeAsO, FeSe and FeTe resolved by the MLWF 
in the $d$ model. 
Energy is measured from the Fermi level. 
Pseudogaps seen in $yz/zx$ and $x^2-y^2$ bands for LaFeAsO 
around $E\sim 0$-0.5 eV are filled in FeTe. 
(Right panel) Partial density of states without interorbital hybridization is 
plotted for comparison. } 
\label{fig:pdos} 
\end{figure*}

Looking at the density of states, 
we find that 
the $d$ band in the 1111 family has a dip (pseudogap) at the energy 
roughly 0.5 eV higher than the Fermi level. 
In other families, the pseudogap is not clear, 
and it is completely smeared out in FeTe. 
The partial density of states and the occupation number
resolved by the MLWF in the $d$ model 
are shown in Fig. \ref{fig:pdos} and Table \ref{tab:number}, respectively, 
for LaFeAsO, FeSe, and FeTe. 
The occupation number is the largest for the $d_{3z^2-r^2}$ orbital which 
is nearly $\frac{3}{4}$-filling in all the compounds. 
Hereafter we abbreviate the 3$d$ orbitals such as $d_{3z^2-r^2}$ as $3z^2-r^2$,
unless confusions occur. 
The ${x^2-y^2}$ and ${yz}/{zx}$ orbitals are roughly half-filling, 
although the weight of the former increases as 
we move from LaFeAsO to FeSe and FeTe. 
It turns out that the $x^2-y^2$ and $yz/zx$ orbitals are 
the primary origin of the evolution of this pseudogap structure.
The pDOS for these orbitals are strongly material dependent. 
The $x^2-y^2$ DOS has a peak at 0.3 eV in FeTe. 
In FeSe, the peak is shifted to 0.7-0.8 eV and 
the pseudogap is formed. 
The peak position further shifts to 1.1 eV in LaFeAsO. 
The $yz/zx$ orbitals show the similar trend.
On the other hand, the lower peaks around $-$0.2 eV are rather pinned through the variation 
from FeTe to LaFeAsO both for $x^2-y^2$ and $yz/zx$ orbitals.  This leaves a pseudogap
progressively in FeSe and then more prominently in LaFeAsO.
The origin of this pseudogap is basically ascribed to the evolution of the $h$ parameter
shown in Table~\ref{tab:geometry}. 
When $h$ decreases, the hybridization between the Fe-3$d$ and pnictogen-/chalcogen-$p$ orbitals becomes 
appreciable and the interorbital hoppings are enhanced, thus generating the splitting of the Fe-3$d$ bands 
into lower and upper bands. 
In the right panel of Fig.\ref{fig:pdos}, the pDOS without interorbital hoppings is plotted. 
Comparison with the left panel clearly shows a crucial role of hybridization effect in forming the pseudogap.
 (we will discuss this pseudogap formation in more detail 
in Sec. \ref{Summary}). 
 In the 1111 family, this large splitting appears to make almost a band insulator with the Fermi level 
 sitting in the ``gap" region between the two split bands for the $x^2-y^2$ and $yz/zx$ orbitals. 
 Because of the incomplete band splitting, the $yz/zx$ as well as $x^2-y^2$ orbitals contribute to 
 the formation of the electron and hole pockets at the Fermi level.  
 This semimetallic nature with small carrier density determines the relatively weakly correlated character of 
the 1111 family, together with the weaker effective Coulomb interaction revealed below. 
The other orbitals, $3z^2-r^2$ and $xy$, have clear gap and do not contribute to the low-energy excitations in all the families. 
 In the 11 family, especially in FeTe, $x^2-y^2$ and $yz/zx$ bands have overall peak at the Fermi level even in 
 each partial DOS and form partially filled bands, in the LDA picture. This large DOS at the Fermi level 
 in the LDA may efficiently trigger the electron correlation effects when we consider the Coulomb interaction beyond the LDA.  
 Indeed, it will be revealed that in the 11 family the effective Coulomb interaction itself is larger than that in the 1111 family.  
 Combination of these two may lead to appreciable electron correlation effects 
 when we go beyond the LDA as is known in the formation of the Mott insulator in the half-filled band of the Hubbard model.  
 We stress again that the evolution from the relatively weak correlation for the 1111 family 
 to the strong correlation for the 11 family emerges only for the $x^2-y^2$ and $yz/zx$ orbitals in an orbital-selective fashion. 
While the combination of the $3z^2-r^2$ and $xy$ 
orbitals is always ``band-insulating" like and does not join in this physics
directly, it should be noted that the formation of the pseudogap
in $x^2-y^2$ and $yz/zx$ is caused by the hybridization
between those orbitals and $3z^2-r^2$ and $xy$, so that it is not
so trivial to derive a three-orbital model for $x^2-y^2$ 
and $yz/zx$.
\begin{table}[htbp] 
\caption{Occupation number of the MLWFs in the $d$ model. 
The number is normalized so that unity corresponds to half filling, 
and thus the sum of the occupancy is 6. 
The occupations of $yz/zx$ and $x^2-y^2$ are roughly close to half filling.}
\  
\label{tab:number} 
\begin{tabular}{lccccc} 
\hline 
\hline
 & $xy$ & $yz$ & $3z^2-r^2$ & $zx$ & $x^2-y^2$  \\ 
\hline 
LaFeAsO & 1.14 & 1.19 & 1.50 & 1.18 & 0.99 \\ 
FeSe & 1.22 & 1.06 & 1.55 & 1.06 & 1.11 \\ 
FeTe & 1.28 & 0.98 & 1.53 & 0.98 & 1.24  \\ 
\hline 
\hline
\end{tabular} 
\end{table} 

\subsection{Wannier functions}
In the present work, we derive two models for each material.  
One is the $d$ model which contains the Fe-3$d$ manifold only. 
The other is the $dp$ model which contains  
the P-3$p$/As-4$p$/Se-4$p$/Te-5$p$ states 
 in addition to the Fe-3$d$ bands.  
We include O-2$p$ states as well in LaFePO and LaFeAsO, 
and derive the $dpp$ model instead of the $dp$ model,  
since they overlap with pnictogen-$p$ band.  
 
\begin{figure}[h!] 
\begin{center} 
\includegraphics[width=0.5\textwidth]{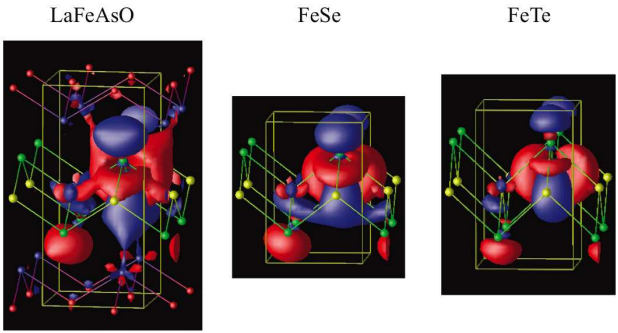}
\end{center} 
\caption{Color:
Isosurface of 
the maximally localized Wannier function at $\pm$0.02 bohr$^{-3/2}$
for the Fe ${x^2-y^2}$ orbital in the $d$ model of 
LaFeAsO (left), FeSe (middle), and FeTe (right). 
This illustrates how the Wannier spread shrinks from LaFeAsO to FeTe.
 The dark shaded surfaces (color in blue) indicate the positive
isosurface at +0.02 and the light shaded surfaces (color in red)
indicate $-$0.02.
}
\label{fig:mlwf_d} 
\end{figure}

\begin{table}[htbp] 
\caption{Spread of the MLWFs (in $\AA^2$)
defined by quadratic extent. 
The spread is the largest for the $x^2-y^2$ orbital except for FeTe.  
} 
\label{tab:spread} 
\begin{tabular}{l@{\,}ccccc} 
\hline 
\hline 
 & $xy$ & $yz$ & $3z^2-r^2$ & $zx$ & $x^2-y^2$  \\ 
\hline 
LaFePO ($d$ model)  & 2.90 & 3.87 & 3.19 & 3.87 & 6.24 \\ 
LaFeAsO ($d$ model) & 2.75 & 3.91 & 3.14 & 3.91 & 5.37 \\ 
BaFe$_2$As$_2$ ($d$ model) & 2.68 & 4.07 & 2.65 & 4.07 & 4.22 \\ 
LiFeAs ($d$ model) & 2.68 & 3.16 & 2.52 & 3.16 & 3.51 \\ 
FeSe ($d$ model) & 1.82 & 2.20 & 1.73 & 2.20 & 2.60 \\ 
FeTe ($d$ model) & 1.98 & 3.51 & 1.94 & 3.51 & 2.48  \\ 
\hline 
LaFePO ($dpp$ model)  & 1.01 & 1.67 & 1.31 & 1.67 & 2.03 \\ 
LaFeAsO ($dpp$ model) & 1.02 & 1.65 & 1.23 & 1.65 & 1.69 \\ 
BaFe$_2$As$_2$ ($dp$ model) & 0.98 & 1.27 & 0.95 & 1.27 & 1.32 \\ 
LiFeAs ($dp$ model) & 1.01 & 1.17 & 0.97 & 1.17 & 1.07 \\ 
FeSe ($dp$ model) & 0.78 & 0.86 & 0.77 & 0.86 & 0.86 \\ 
FeTe ($dp$ model) & 0.94 & 0.97 & 0.90 & 0.97 & 0.86 \\ 
\hline 
\hline
\end{tabular} 
\end{table} 
Table \ref{tab:spread} shows the spread (quadratic extent) of the MLWFs.  
Note that the $xy$ axes in our convention are along the unit vectors of 
the cell containing two Fe atoms (Fig.~\ref{fig:geometry}), 
while they are sometimes rotated by 45 degrees in other works. 
One striking feature is that the 
Wannier orbitals are strongly orbital dependent in the $d$ model.  
The $x^2-y^2$ orbital is the most extended except for FeTe,  
whereas the $xy$ and $3z^2-r^2$ are localized.  
The anisotropy is enhanced for the 1111 family,  
for which the absolute values of the spread are large.  
These trends are understood as follows.  
The ten states in the $d$ model contain considerable  
pnictogen-/chalcogen-$p$ component.  
The hybridization between the $p$ and Fe-$d$ states make 
the Wannier functions delocalized.  
The hybridization becomes stronger in the ${x^2-y^2}$ Wannier orbital,  
because the orbital is directed to the pnictogen/chalcogen atoms, 
consequently the spread increases.  
Concerning the family dependence, 
pnictogen/chalcogen atoms approach the Fe plane 
in the order of the 11 $\rightarrow$ 122 $\rightarrow$ 1111  
families (see Table~\ref{tab:geometry}). 
Accordingly the 1111 family has larger hybridization effects.  
This trend is confirmed by comparing the Wannier orbitals 
(Fig.~\ref{fig:mlwf_d}), 
where we can see that the Wannier orbital is more extended in LaFeAsO than in FeTe. 

\begin{figure}[htbp] 
\begin{center} 
\includegraphics[width=0.5\textwidth]{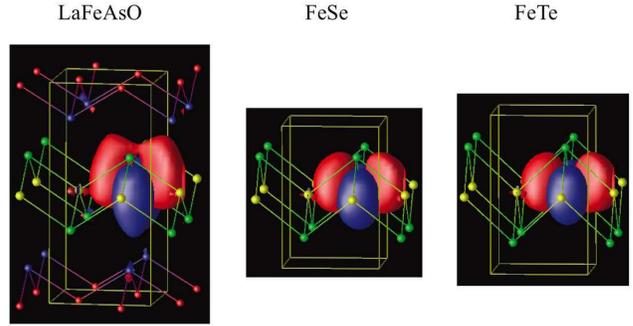}
\end{center} 
\caption{Color:
Isosurface of 
the maximally localized Wannier function at $\pm$ 0.02 bohr$^{-3/2}$
for the Fe-$d_{x^2-y^2}$ orbital
in the $dpp/dp$ model of
LaFeAsO (left), FeSe (middle), and FeTe (right). 
The dark shaded surfaces (color in blue) indicate the positive
isosurface at +0.02 and the light shaded surfaces (color in red)
indicate $-$0.02.
} 
\label{fig:mlwf_dp} 
\end{figure}

Comparing the $dp/dpp$ model with the $d$ model,  
we find that
the former has smaller spread 
as seen in comparison of Figs.~\ref{fig:mlwf_d} and \ref{fig:mlwf_dp}; 
In the $dp/dpp$ model, 
the Wannier functions are constructed from a larger number of the 
Kohn-Sham states,  and therefore, the optimized Wannier orbitals are more localized 
 and contain less $p$ character (see Table~\ref{tab:spread}). 
They are more atomic-orbital like and the orbital dependence is weaker.

\subsection{Transfer integrals of $d$ model}
Tables \ref{t_1111}-\ref{t_11} show the transfer integrals, 
$t_{mn}({\bR})$, in the $d$ model. 
(In the tables and this subsection, 
the symmetry of the $d$ orbitals is denoted  
as the number; 1 for ${xy}$, 2 for ${yz}$, 3 for ${3z^{2}-r^2}$, 4 for ${zx}$, and 5 for ${x^2-y^2}$ orbitals.) 
The hopping parameters between the neighboring Fe sites are  
listed in the column for $(R_x,R_y,R_z) = (0.5,-0.5,0)$.  
The largest component is for 
($m$, $n$)=($xy$, $xy$), $t_{11}$. 
Its value is 0.3-0.4 eV and is slightly larger in the 11 and 111 families  
than in the 1111 and 122 families, 
in accordance with shorter in-plane lattice constant.  
The next nearest hopping $(R_x,R_y,R_z) = (1,0,0)$  
is comparable with the nearest neighbor hopping. 
Especially, the ($m$, $n$)=($zx$, $zx$) component, $t'_{44}$,  
is larger than $t_{11}$ in the 1111 and 122 families. 
The ratio $t'_{44} / t_{11}$, which can be regarded as a measure of the frustration, 
is larger in the 1111 family than in the 11 family. 
Other transfer integrals associated with the $yz$ or $zx$ 
orbital are also large. 
These orbitals (in the $d$ model) are well extended, 
hence hopping through the pnictogen/chalcogen atom would 
contribute to the transfer integral. 
Longer-range hoppings are also nonnegligible. 
In fact, if we neglect transfer integrals with less than 0.05 eV in the absolute value, 
the band width is reduced by 12~\% (7~\%) in LaFeAsO (FeSe).
The transfer integrals in the $c$ direction are sizable in BaFe$_2$As$_2$, 
particularly for the $x^2-y^2$ orbital, 
indicating three-dimensional character of the electronic states. 
They are much smaller in the 1111 family. 

\begin{table*}[htb] 
\caption{Transfer integrals in the $d$ model, $t_{mn}(R_x, R_y, R_z)$, 
where $m$ and $n$ specify symmetry of $d$ orbitals; 1 for ${xy}$, 2 for ${yz}$, 3 for ${3z^{2}-r^2}$, 4 for ${zx}$, 
and 5 for ${x^2-y^2}$ orbitals. Symmetry operations of 
$\sigma_{y}$, $I$, and $\sigma_{d}$ change $t_{mn}(R_x, R_y, R_z)$ to $t_{mn}(R_x, -R_y, R_z)$, $t_{mn}(-R_x,-R_y,R_z)$, 
and $t_{mn}(R_y,R_x,R_z)$. Notice also that $t_{mn}(\bR)=t_{nm}(-\bR)$. 
Since the calculations are carried out using the 4$\times$4$\times$4 $k$-mesh,
the Wannier functions are periodic with the period of (4,0,0).
Because of this periodicity, the transfer integrals for ${\bf R}$=[2,0,0]
defined by eq.(1) are twice compared to the case
in which the period of the Wannier functions is sufficiently long.
We therefore halved the values for ${\bf R}$=[2,0,0] in the tables,
which are more appropriate as parameters for the model Hamiltonian.
Units are given in meV. 
} 
\
{\scriptsize 
\begin{tabular}{c|rrrrrrrrrrc} 
\hline \hline \\ [-4pt]
LaFePO \\ [+2pt] 
\hline \\ [-4pt]
\backslashbox{$(m, n)$}{$\bR$} 
& \big[0,0,0\big] 
& \big[$\frac{1}{2}$,$-\frac{1}{2}$,0\big] 
& \big[1,0,0\big] 
& \big[1,$-$1,0\big] 
& \big[$\frac{3}{2}$,$-\frac{1}{2}$,0\big] 
& \big[2,0,0\big] 
& \big[0,0,$\frac{c}{a}$\big] 
& \big[$\frac{1}{2}, -\frac{1}{2}, \frac{c}{a}$\big] 
& $\sigma_{y}$ 
& $I$ 
& $\sigma_{d}$ \\ [+4pt]
\hline \\ [-8pt]
$(1,1)$& 598&$-$342&$-$106&$-$25&$-$11&$-$7&$-$1&   1& + &  +  &   +    \\
$(1,2)$&   0&   307&   172&   12&   18&    0&  0 &   0& + & $-$ &$-$(1,4)\\
$(1,3)$&   0&$-$327&     0&$-$21&$-$29&    0&  0 &$-$1&$-$&  +  &   +    \\
$(1,4)$&   0&   307&     0&   12&   45&    0&  0 &$-$1&$-$& $-$ &$-$(1,2)\\
$(1,5)$&   0&     0&     0&    0&   24&    0&  0 &$-$3&$-$&  +  &  $-$   \\
$(2,2)$& 886&   231&   141&   28&   15&   12&  3 &   6& + &  +  & (4,4)  \\
$(2,3)$&   0& $-$31&     0&$-$13& $-$8&    0&  0 &   1&$-$& $-$ &$-$(4,3)\\
$(2,4)$&   0&   134&     0&$-$15& $-$2&    0&  0 &   0&$-$&  +  & (4,2)  \\
$(2,5)$&   0&   164&     0& $-$7&   26&    0&  0 &   3&$-$& $-$ & (4,5)  \\
$(3,3)$& 684&   130&    22&$-$42&$-$23&$-$20&$-$7&$-$2& + &  +  &   +    \\
$(3,4)$&   0&    31&   115&   13&   10&    0&  0 &$-$1& + & $-$ &$-$(3,2)\\
$(3,5)$&   0&     0&$-$234&    0&    4&   11&  2 &$-$2& + &  +  &  $-$   \\
$(4,4)$& 886&   231&   395&   28&   47&  82&  3 &   1& + &  +  &  (2,2) \\
$(4,5)$&   0&$-$164& $-$77&    7&$-$17&    0&  0 &   2& + & $-$ &  (2,5) \\
$(5,5)$&1234&$-$257&   156&$-$56&   33&$-$27&$-$2&   1& + &  +  &   +    \\
\hline \hline \\ [-4pt]

LaFeAsO \\ [+2pt] 
\hline \\ [-4pt]
\backslashbox{$(m, n)$}{$\bR$} 
& \big[0,0,0\big] 
& \big[$\frac{1}{2}$,$-\frac{1}{2}$,0\big] 
& \big[1,0,0\big] 
& \big[1,$-$1,0\big] 
& \big[$\frac{3}{2}$,$-\frac{1}{2}$,0\big] 
& \big[2,0,0\big] 
& \big[0,0,$\frac{c}{a}$\big] 
& \big[$\frac{1}{2}, -\frac{1}{2}, \frac{c}{a}$\big] 
& $\sigma_{y}$ 
& $I$ 
& $\sigma_{d}$ \\ [+4pt]
\hline \\ [-8pt]
$(1,1)$&790 & $-$315 &$-$67 & $-$19& $-$2 &  1 & $-$2&1 & +  &  +  &   + \\
$(1,2)$&  0 & 253 &138  & 1  & 10 &  0   &  0  &   0 & +  & $-$ & $-$(1,4)  \\
$(1,3)$&  0 & $-$301 & 0  &1  & $-$18 &  0   &  0  &   0 & $-$ &  +  &   +   \\
$(1,4)$&  0 & 253 & 0  & 1  &33 &  0   &  0  &   $-$1 & $-$ & $-$ &$-$(1,2) \\
$(1,5)$&  0 & 0 & 0  &    0  & 10 &  0&  0  &   $-$2 & $-$ &  +  &    $-$    \\
$(2,2)$& 1099 & 206 &135 & 12  & 9 &  5   &  1  &   7 &  +  &  +  & (4,4)   \\
$(2,3)$&  0 & $-$73 &0  & $-$2& $-$1 &  0 &  0  &  2 & $-$ & $-$ &$-$(4,3) \\  
$(2,4)$&  0 & 137 & 0  & $-$18& $-$9 &  0   &  0  &   1 & $-$ &  +  & (4,2) \\
$(2,5)$&  0 & 165 &   0  & -4  & 10 &  0   &  0  &   3 & $-$ & $-$ & (4,5)  \\
$(3,3)$&890 &72 & $-$13& $-$38  & $-$15 & $-$18 &$-$6 &$-$2 &  +  &  +  &   +\\
$(3,4)$&  0 & 73 & 137 & 2  & $-$3 &  0 &  0  & $-$1 &  +  & $-$ & $-$(3,2)  \\
$(3,5)$&  0 & 0 & $-$159 & 0  & 1 &  17 &  3  &   $-$3 &  +  &  +  &   $-$   \\
$(4,4)$& 1099 &206 & 345& 12  & 36 &70 &  1  &   0 &  +  &  +  &  (2,2)  \\
$(4,5)$&  0 & $-$165 & 19 & 4 & $-$11 &  0 &  0  &   1 &  +  & $-$ & (2,5) \\
$(5,5)$& 1255 & $-$152 &118& $-$24  & 30 & $-$28 &  1 &$-$2 &  +  &  +  & + \\
\hline 
\hline 
\end{tabular}
}
\label{t_1111} 
\end{table*} 

\begin{table*}[htb] 
\caption{Transfer integrals in the $d$ model for BaFe$_2$As$_2$. 
Notations are the same as Table~\ref{t_1111}. 
} 
\ 
{\scriptsize 
\begin{tabular}{c|rrrrrrrrc}
\hline \hline \\ [-4pt]
BaFe$_2$As$_2$ \\ [+2pt] 
\hline \\ [-4pt]
\backslashbox{$(m, n)$}{$\bR$} 
& \big[0,0,0\big] 
& \big[$\frac{1}{2}$,$-\frac{1}{2}$,0\big] 
& \big[1,0,0\big] 
& \big[1,$-$1,0\big] 
& \big[$\frac{3}{2}$,$-\frac{1}{2}$,0\big] 
& \big[2,0,0\big] 
& $\sigma_{y}$ 
& $I$ 
& $\sigma_{d}$ \\ [+4pt]
\hline \\ [-8pt]
$(1,1)$   &$-$127&$-$341  & $-$66 & $-$18  & $-$1 & 10  & +  &  +  &   +     \\
$(1,2)$   &  0 &    260 &   134  & $-$4  & $-$3 & 0  & +  & $-$ &$-$(1,4)  \\
$(1,3)$   &  0 & $-$314 &   3  &    14  & $-$16 & $-$2  &$-$ &  +  & +   \\
$(1,4)$   &  0 &    255 &   3  & $-$7  &    32 & 0  &$-$ & $-$ &$-$(1,2) \\
$(1,5)$   & $-$8 &  1 &   1  &    1  & $-$2 &  0  &$-$ &  +  & $-$    \\
$(2,2)$   &  267 &    209 &   130  &    2  & 3 & 6 & +  &  +  & (4,4)   \\
$(2,3)$   &  0 & $-$89 &   2  &    1  & $-$2 &  0   &$-$ & $-$ &$-$(4,3) \\
$(2,4)$   &  0 &    123 &   2  & $-$27  & $-$18 & $-$1 &$-$ &  +  & (4,2) \\
$(2,5)$   &  0 &    183 & $-$3 & $-$6  &    11 & 0   &$-$ & $-$ &  (4,5)  \\
$(3,3)$   &  5 &    45 & $-$16 & $-$38  & $-$18 & $-$15 & +  &  +  &   +     \\
$(3,4)$   &  0 &    89 &   158  & $-$4  & $-$5 & 0   & +  & $-$ & $-$(3,2)  \\
$(3,5)$   &  0 &    0 & $-$160 &    0  &   6 &  17   & +  &  +  &   $-$   \\
$(4,4)$   &  267 &    210 &   356  & 1&41 &  73   & +  &  +  &  (2,2)  \\
$(4,5)$   &  0 & $-$178 &    55 &    10  & $-$11 & 0   & +  & $-$ & (2,5) \\
$(5,5)$   & 363  & $-$133 &    107 & $-$30  & 41 & $-$28 & +  &  +  &    +    \\
\hline \\ [-4pt]
\backslashbox{$(m, n)$}{$\bR$} 
& \big[$    0,             0,    \frac{c}{2a}$\big]$_{{\rm AB}}$  
& \big[$    0,             0    ,\frac{c}{2a}$\big]$_{{\rm BA}}$  
& \big[$ \frac{ 1}{2}, \frac{ 1}{2}, \frac{ c}{2a}$\big] 
& \big[$ \frac{ 1}{2}, \frac{ 1}{2}, -\frac{ c}{2a}$\big] 
 \\ [+4pt]
\hline \\ [-8pt]
$(1,1)$   & $-$10 & $-$10 & $-$4 & $-$4 \\
$(1,2)$   &    0 &    0 &    3 & $-$16 \\
$(1,3)$   &    1 &    1 &    2 &  2 \\
$(1,4)$   &    0 &    0 &    16 &   $-$3 \\
$(1,5)$   &    0 &    0 &    0 &    0 \\
$(2,2)$   & $-$23 & $-$23 & $-$14 & $-$14 \\
$(2,3)$   &    0 &    0 &    1 &    $-$7 \\
$(2,4)$   &    0 &    0 & $-$11 &    1 \\
$(2,5)$   &    0 &    0 & $-$10 & $-$1 \\
$(3,3)$   & $-$86 & $-$86 & $-$37 & $-$37 \\
$(3,4)$   &    0 &    0 &    1 & $-$7 \\
$(3,5)$   & $-$111 &    111 &    51 & $-$51 \\
$(4,4)$   & $-$23 & $-$23 & $-$14 & $-$14 \\
$(4,5)$   &    0 &    0 &    1 & $-$10 \\
$(5,5)$   & $-$162 & $-$162 &    85 &  85 \\
\hline 
\hline 
\end{tabular}
} 
\label{t_122} 
\end{table*} 
\begin{table*}[htb] 
\caption{Transfer integrals in the $d$ model for LiFeAs. 
Notations are the same as Table~\ref{t_1111}. 
} 
\ 
{\scriptsize 
\begin{tabular}{c|rrrrrrrrrrc} 
\hline \hline \\ [-4pt]
LiFeAs \\ [+2pt] 
\hline \\ [-4pt]
\backslashbox{$(m, n)$}{$\bR$} 
& \big[0,0,0\big] 
& \big[$\frac{1}{2}$,$-\frac{1}{2}$,0\big] 
& \big[1,0,0\big] 
& \big[1,$-$1,0\big] 
& \big[$\frac{3}{2}$,$-\frac{1}{2}$,0\big] 
& \big[2,0,0\big] 
& \big[0,0,$\frac{c}{a}$\big] 
& \big[$\frac{1}{2}, -\frac{1}{2}, \frac{c}{a}$\big] 
& $\sigma_{y}$ 
& $I$ 
& $\sigma_{d}$ \\ [+4pt]
\hline \\ [-8pt]
$(1,1)$&$-$180& $-$409 &$-$26 & $-$57& 1 &25  &$-$32 &9 & +  &  +  &   +     \\
$(1,2)$&  0 &290 &144& $-$3  & $-$15 &0  &0  & $-$10 & +  & $-$ &$-$(1,4)  \\
$(1,3)$ &  0 & $-$346 &   0  &    35  & $-$8 & 0  &0  & 17 &$-$ &  +  & +   \\
$(1,4)$ &  0 &290 &   0  & $-$3  & 56 & 0  & 0  & $-$3 &$-$ & $-$ &$-$(1,2) \\
$(1,5)$&  0 & 0 &   0  &    0  & $-$19 & 0  & 0  & $-$7 &$-$ &  +  & $-$    \\
$(2,2)$&  270 & 229 &142& $-$32& $-$21 & $-$9 & 11&39 & +  &  +  & (4,4)   \\
$(2,3)$&  0 & $-$127 & 0  & 11  & 4 &  0   & 0  & 6 &$-$ & $-$ &$-$(4,3) \\
$(2,4)$&  0 &167 &0  & $-$42  & $-$30 & 0   & 0  &15 &$-$ &  +  & (4,2) \\
$(2,5)$&  0 &198 &   0  & $-$9  & $-$3 & 0   & 0  & 18 &$-$ & $-$ &  (4,5)  \\
$(3,3)$& $-$133 &$-$26& $-$52 & $-$13&5 &$-$19 &$-$65& $-$31 & +  &  +  & + \\
$(3,4)$&  0 &127 &188& $-$11& $-$18 & 0&  0& $-$35 & +& $-$ & $-$(3,2)  \\
$(3,5)$&  0 & 0 & $-$79 &0 & $-$5 & 17   & 46 &  $-$42 & +  &  +  &   $-$   \\
$(4,4)$&271 & 230 & 414& $-$32  & 52 &  93   & 11  & 5 & +  &  +  &  (2,2)  \\
$(4,5)$&  0 & $-$198 &124 &  9  & $-$19 & 0   & 0  & 30 & +  & $-$ & (2,5) \\
$(5,5)$& 137  & $-$7 &81 &1& 35 & $-$37 & 58  & $-$51 & +  &  +  &    +    \\
\hline 
\hline 
\end{tabular}
} 
\label{t_111} 
\end{table*} 
\begin{table*}[htb] 
\caption{Transfer integrals in the $d$ model for FeSe and FeTe.
Notations are the same as Table~\ref{t_1111}. 
} 
\ 
{\scriptsize 
\begin{tabular}{c|rrrrrrrrrrc} 
\hline \hline \\ [-4pt]
FeSe \\ [+2pt] 
\hline \\ [-4pt]
\backslashbox{$(m, n)$}{$\bR$} 
& \big[0,0,0\big] 
& \big[$\frac{1}{2}$,$-\frac{1}{2}$,0\big] 
& \big[1,0,0\big] 
& \big[1,$-$1,0\big] 
& \big[$\frac{3}{2}$,$-\frac{1}{2}$,0\big] 
& \big[2,0,0\big] 
& \big[0,0,$\frac{c}{a}$\big] 
& \big[$\frac{1}{2}, -\frac{1}{2}, \frac{c}{a}$\big] 
&   $\sigma_{y}$ 
& $I$ 
& $\sigma_{d}$ \\ [+4pt]
\hline \\ [-8pt]
$(1,1)$&854 & $-$410 & $-$69 & $-$10  & 4 &11&$-$24 &7 & +  &  +  &   +  \\
$(1,2)$& 0 &272 &131  & $-$9  & $-$6 &0  &  0 &  $-$8 & +  & $-$ &$-$(1,4)  \\
$(1,3)$   & 0 & $-$347 & 0  & 22  & $-$7 &0  & 0  &11 &$-$ &  +  & +   \\
$(1,4)$   &0 &    272 & 0  & $-$9  & 23 &0  &0  & $-$2 &$-$ & $-$ &$-$(1,2) \\
$(1,5)$   &0 & 0 & 0  & 0  & $-$8 &0  & 0  &  $-$4 &$-$ &  +  & $-$    \\
$(2,2)$&1418 &199 &126  & $-$16  & $-$8 & $-$6 &9  &30 & +  &  +  & (4,4)   \\
$(2,3)$   & 0 & $-$120 & 0  & 7  & 4 &  0& 0  &10 &$-$ & $-$ &$-$(4,3) \\
$(2,4)$   & 0 & 127 &0  & $-$24  & $-$18 &  0& 0  &11 &$-$ &  +  & (4,2) \\
$(2,5)$   &0 &    223 & 0  & 0  & $-$4 &  0  & 0  &20 &$-$ & $-$ &  (4,5)  \\
$(3,3)$ &980 & $-$3 & $-$18 & $-$16  & $-$3 & $-$15 &$-$22 & $-$8 & +&+  & + \\
$(3,4)$& 0 & 120 &196& $-$7& $-$12 &  0& 0  &  $-$10 & +  & $-$ & $-$(3,2)  \\
$(3,5)$   &0 & 0 & $-$115 & 0  & $-$3 & 10& $-$7&  $-$6 & +  &  +  &   $-$   \\
$(4,4)$   &  1418 &199 & 348  & $-$16  &15 &59& 9  & 1 & +  &  +  &  (2,2)  \\
$(4,5)$   & 0 & $-$223 & 82 & 0  & $-$13 &0   &  0& 7 & +  & $-$ & (2,5) \\
$(5,5)$   & 1335 & $-$57 & 92 & $-$1  & 19 & $-$24 & $-$29&  5 & +  &  +  &    +    \\
\hline \hline \\ [-4pt]

FeTe \\ [+2pt] 
\hline \\ [-4pt]
\backslashbox{$(m, n)$}{$\bR$} 
& \big[0,0,0\big] 
& \big[$\frac{1}{2}$,$-\frac{1}{2}$,0\big] 
& \big[1,0,0\big] 
& \big[1,$-$1,0\big] 
& \big[$\frac{3}{2}$,$-\frac{1}{2}$,0\big] 
& \big[2,0,0\big] 
& \big[0,0,$\frac{c}{a}$\big] 
& \big[$\frac{1}{2}, -\frac{1}{2}, \frac{c}{a}$\big] 
& $\sigma_{y}$ 
& $I$ 
& $\sigma_{d}$ \\ [+4pt]
\hline \\ [-8pt]
$(1,1)$&163 & $-$392 & $-$2 & $-$14  &2 &8& $-$39 & 10 & +  &  +  &   +  \\
$(1,2)$& 0 &228 & 96& $-$9  & $-$13 &0   &  0& $-$15 & +  & $-$ & $-$(1,4)  \\
$(1,3)$   & 0 & $-$341 & 0  & 33  & $-$6 & 0   &0   & 21 &$-$ &  +  &   +   \\
$(1,4)$   &0 & 228 & 0  & $-$9  &39 &0   &  0& 1 &$-$ & $-$ &$-$(1,2) \\
$(1,5)$   & 0 & 0 &0  & 0  & $-$15 &0   &0   & 0 &$-$ &  +  & $-$    \\
$(2,2)$& 774 & 164& 99  &$-$33 & $-$17 & $-$15 & 10& 37 & +  &  +  & (4,4)   \\
$(2,3)$   & 0 & $-$130 &0  & 16  & 8 &  0 &0   & 15 &$-$ & $-$ &$-$(4,3) \\
$(2,4)$   & 0 &107 & 0  & $-$31  & $-$34 & 0   &  0 & 17 &$-$ &  +  & (4,2) \\
$(2,5)$   & 0 & 184 & 0  & $-$5& $-$9 & 0   &  0 & 18 &$-$ & $-$ & (4,5)  \\
$(3,3)$&189& $-$87 & $-$68 & 4& 19 & $-$18 &$-$47& $-$12 & +  &  +  &   +  \\
$(3,4)$& 0 & 130 & 200 & $-$16& $-$11 &  0 &  0 & $-$30 & +  & $-$ & $-$(3,2)\\
$(3,5)$&0 & 0 & $-$23 &0  & $-$13 &13   & 15 & $-$20 & +  &  +  &   $-$   \\
$(4,4)$& 774 & 164  & 348 &  $-$32 &47 &  80 &10   &11 & +  &  +  &  (2,2)  \\
$(4,5)$   & 0 & $-$184 & 144 &5  & $-$26 & 0 &  0&25 & +  & $-$ & (2,5) \\
$(5,5)$   & 466 & 85 & 40 &9 & 10 & $-$16 &17  &  $-$25 & +  &  +  &    +    \\
\hline 
\hline 
\end{tabular}
} 
\label{t_11} 
\end{table*} 

Here we remark the rule to derive the transfers not explicitly shown in Tables \ref{t_1111}-\ref{t_11} 
 by the symmetry operations.
As is shown in Fig.~\ref{fig:geometry}, there are two iron atoms (Fe-A and Fe-B)
in the unit cell. Note that we can choose arbitrarily the phase
of the transfer hoppings between 
Fe-A and Fe-B sites [i.e., $t_{mAnB}({\bR})$]
by introducing a local gauge to one of these irons.
Let us first look at the case of $R_z=0$.
If we employ a common (global) coordinate for the irons
and define a common phase of 
$\phi_{nA{{\bR}}}$ and $\phi_{nB{{\bR}}}$  
(as was done in ref.~\citen{Nakamura}),
then $t_{mAnB}({\bR}) = s \times t_{mBnA}({\bR})$ with $s$=$-1$ when
one and only one of $m$ and $n$ is ${yz}$ or ${zx}$ orbitals. Otherwise $s=1$. 
However, as was done in ref.~\citen{Kuroki} if we attach
a gauge field $\exp(i \pi)$ to $\phi_{{yz}B{{\bf R}}}$ and 
$\phi_{{zx}B{{\bf R}}}$, then
$t_{mAnB}({\bR})$ becomes equal to $t_{mBnA}({\bR})$ for all $m$ and $n$.

In addition, the transfer between two irons on the same sublattice $t_{mAnA}({\bR})=t_{mBnB}({\bR})$ 
also has translational symmetry irrespective of the choice of this sublattice dependent gauge. 
Thus, if we choose the above appropriate gauge,
the transfer integrals in these systems depend only on $m$ and $n$ 
and ${\bR}$ irrespective of the sublattice of iron, which allows an apparent higher translational symmetry.
The transfer constructed in this gauge gains an apparent square lattice symmetry without distinction of the two sublattice points. 

For the cases of 1111, 111, and 11 where the translation of $(0,0,l)$ ($l$, any integer) generates 
the layer structure at $R_z \neq 0$ identical to that at $R_z=0$ shown in Fig.~\ref{fig:geometry} 
in terms of the pnictogen/chalcogen positions relative to the iron layers, 
whereas, in the case of 122, the identical layer structure is obtained only after 
$(\frac{1}{2},\frac{1}{2},0)$ translation, because of the antiphase of the As positions.
In other words, the conventional cell of 122 contains two iron layers because of the body center tetragonal symmetry.

Pnictogen/chalcogen 
positions are in-phase along the $c$ axis 
for the 1111, 111 and 11 families 
, this higher translational symmetry holds
for general $R_z$ for the present local gauge $\exp(i \pi)$ to $\phi_{{yz}B{\bR}}$ and
$\phi_{{zx}B{\bR}}$. 
Therefore, we can unfold the Brillouin zone
(BZ). This is convenient when we want to solve the obtained effective lattice model numerically because
the unit cell is halved with only one iron site contained. 
In Tables \ref{t_1111}-\ref{t_11}, we list the transfer integrals for this gauge.
Although the five-band model in this extended BZ is generally more
convenient than the ten-band model in the original BZ
for numerical model  calculations, note that we have to fold the Brillouin
zone again to compare the result with experiments.

For 122, while the relation
$t_{mAnB}({\bR}) = t_{mBnA}({\bR})$ holds
for $R_z=0$, it is not satisfied for $R_z \neq 0$ because of the above mentioned antiphase of the As position.
Namely, the BZ can not
be unfolded. For example, $t_{mAnB} (0,0,-\frac{c}{2a})$ is not necessarily equal 
to
$t_{mBnA}(0,0,-\frac{c}{2a})$, so that we list both of them in Table \ref{t_122}.

In Tables \ref{t_1111}-\ref{t_11}, we also show that how the transfer hoppings change their 
signs
by symmetry operations. For example, the transfer hoppings
of $t_{mn}(\frac{1}{2},\frac{1}{2},0)$ is obtained by applying $\sigma_y$
to the transfer for ${\bR}$=$(\frac{1}{2},-\frac{1}{2},0)$, 
namely $t_{mn}(\frac{1}{2},\frac{1}{2},0)$=$\sigma_y t_{mn}(\frac{1}{2},-\frac{1}{2},0)$,  
while those of ${\bR}$=$(-\frac{1}{2},\frac{1}{2},0)$ is obtained by applying $I$ instead of $\sigma_y$.
The transfer for ${\bR}$=$(\frac{1}{2},\frac{3}{2},0)$ is obtained by applying 
$\sigma_y$, $I$, and $\sigma_d$ to ${\bR}$=$(\frac{3}{2},-\frac{1}{2},0)$ 
as $t_{mn}(\frac{1}{2},\frac{3}{2},0)=\sigma_yI\sigma_d t_{mn}(\frac{3}{2},-\frac{1}{2},0)$. 
In the column $\sigma_d$, $\pm(m,n)$ means that 
one should take $\pm t_{mn}$ at the same $\bR$ irrespective 
of the column index $(m,n)$.

\subsection{Screened Coulomb interaction of $d$ model 
\label{subsec:U}}
Table \ref{tab:uj_d} shows the onsite screened Coulomb interaction, $U_{mn}(\mathbf{0})$,  
and onsite screened exchange interaction, $J_{mn}(\mathbf{0})$,  
in the $d$ model.  
As has been reported previously for the 1111 family~\cite{Nakamura},  
$U$ depends strongly on the orbital 
by the amount as large as $>$1 eV.  
The orbital dependence is seen in other families as well.  
The values of $U$ decrease and the anisotropy is enhanced 
as the Wannier orbital is extended. 
 \begin{table*} 
\caption{
Effective on-site Coulomb ($U$)/exchange ($J$) interactions between two electrons 
on the same iron site in the $d$ model for all the combinations of iron-3$d$ orbitals (in eV). 
}
\ 
\label{tab:uj_d} 
{\scriptsize
\begin{tabular}{ccccccccccccccc} 
\hline \hline \\ [-8pt]  
LaFePO &      &      & $U$   &      &          &  &     &     &      &  $J$   &      &    \\ [+1pt]
\hline \\ [-8pt] 
 & $xy$ & $yz$ & $3z^2-r^2$ & $zx$ & $x^2-y^2$  &  &  & $xy$ & $yz$ & $3z^2-r^2$ & $zx$ & $x^2-y^2$  \\ 
\hline \\ [-8pt] 
$xy$ & 2.98 & 1.80 & 1.78 & 1.80 & 1.77  & & $xy$ & & 0.45 & 0.54 & 0.45 & 0.20 \\ 
$yz$ & 1.80 & 2.42 & 1.97 & 1.64 & 1.46  & & $yz$ & 0.45 & & 0.32 & 0.36 & 0.31 \\ 
$3z^2-r^2$ & 1.78 & 1.97 & 2.81 & 1.97 & 1.46 & & $3z^2-r^2$ & 0.54 & 0.32 & & 0.32 & 0.37 \\ 
$zx$ & 1.80 & 1.64 & 1.97 & 2.42 & 1.46  & & $zx$ & 0.45 & 0.36 & 0.32 & & 0.31 \\ 
$x^2-y^2$ & 1.77 & 1.46& 1.46 & 1.46 & 1.68 & & $x^2-y^2$ & 0.20 & 0.31 & 0.37 & 0.31 & \\ 
\hline \hline \\ [-8pt]
LaFeAsO &      &      & $U$   &      &          &  &     &     &      &  $J$   &      &    \\ [+1pt]
\hline \\ [-8pt] 
        & $xy$ & $yz$ & $3z^2-r^2$ & $zx$ & $x^2-y^2$ & &     & $xy$ & $yz$ & $3z^2-r^2$ & $zx$ & $x^2-y^2$  \\ 
\hline \\ [-8pt] 
$xy$ & 3.03  & 1.80  & 1.78 & 1.80 & 1.91 &   & $xy$     &      & 0.46 & 0.57 & 0.46 & 0.23 \\ 
$yz$ & 1.80  & 2.43 & 1.97 & 1.62 & 1.52  &   & $yz$     & 0.46 &      & 0.33 & 0.37 & 0.35 \\ 
$3z^2-r^2$ & 1.78  & 1.97 & 2.84 & 1.97 & 1.51 &   & $3z^2-r^2$    & 0.57 & 0.33 &      & 0.33 & 0.42 \\ 
$zx$ & 1.80  & 1.62 & 1.97 & 2.43 & 1.52  &   & $zx$     & 0.46 & 0.37 & 0.33 &      & 0.35 \\ 
$x^2-y^2$ & 1.91  & 1.52 & 1.51 & 1.52 & 1.91 & &$x^2-y^2$ & 0.23 & 0.35 & 0.42 & 0.35 &      \\
\hline 
\hline 
BaFeAs &      &      & $U$   &      &          &  &     &     &      &  $J$   & 2     &    \\ [+1pt]
\hline \\ [-8pt] 
 & $xy$ & $yz$ & $3z^2-r^2$ & $zx$ & $x^2-y^2$  & & & $xy$ & $yz$ & $3z^2-r^2$ & $zx$ & $x^2-y^2$  \\ 
\hline 
$xy$ & 3.18 & 1.94 & 1.99 & 1.94 & 2.16  & & $xy$ & & 0.48 & 0.60 & 0.48 & 0.26 \\ 
$yz$ & 1.94 & 2.64 & 2.21 & 1.77 & 1.72  & & $yz$ & 0.48 & & 0.36 & 0.40 & 0.41 \\ 
$3z^2-r^2$ & 1.99 & 2.21 & 3.28 & 2.21 & 1.77 & & $3z^2-r^2$ & 0.60 & 0.36 & & 0.36 & 0.50 \\ 
$zx$ & 1.94 & 1.77 & 2.21 & 2.64 & 1.72  & & $zx$ & 0.48 & 0.40 & 0.36 & & 0.41 \\ 
$x^2-y^2$ & 2.16 & 1.72 & 1.77 & 1.72 & 2.29  & & $x^2-y^2$ & 0.26 & 0.41 & 0.50 & 0.41 & \\ 
\hline 
\hline 
LiFeAs &      &      & $U$   &      &          &  &     &     &      &  $J$   &      &    \\ [+1pt]
\hline \\ [-8pt] 
    & $xy$ & $yz$ & $3z^2-r^2$ & $zx$ & $x^2-y^2$   &   &  & $xy$ & $yz$ & $3z^2-r^2$ & $zx$ & $x^2-y^2$  \\ 
\hline 
$xy$ & 3.39 & 2.23 & 2.27 & 2.23 & 2.54 & & $xy$ & & 0.47 & 0.60 & 0.47 & 0.28 \\ 
$yz$ & 2.23 & 2.96 & 2.52 & 2.08 & 2.11 & & $yz$ & 0.47 & & 0.36 & 0.39 & 0.44 \\ 
$3z^2-r^2$ & 2.27 & 2.52 & 3.58 & 2.52 & 2.15 & & $3z^2-r^2$ & 0.60 & 0.36 & & 0.36 & 0.54 \\ 
$zx$ & 2.23 & 2.08 & 2.52 & 2.96 & 2.11 & &  $zx$ & 0.47 & 0.39 & 0.36 & & 0.44 \\ 
$x^2-y^2$ & 2.54 & 2.11 & 2.15 & 2.11 & 2.85 & & $x^2-y^2$ & 0.28 & 0.44 & 0.54 & 0.44 & \\ 
\hline 
\hline 
FeSe &      &      & $U$   &      &          &  &     &     &      &  $J$   &      &    \\ [+1pt]
\hline \\ [-8pt] 
  & $xy$ & $yz$ & $3z^2-r^2$ & $zx$ & $x^2-y^2$   &   &  & $xy$ & $yz$ & $3z^2-r^2$ & $zx$ & $x^2-y^2$  \\ 
\hline 
$xy$ & 4.51 & 3.19 & 3.20 & 3.19 & 3.49 & &  $xy$ & & 0.57 & 0.69 & 0.57 & 0.32\\ 
$yz$ & 3.19 & 4.11 & 3.52 & 3.02 & 2.98 & & $yz$ & 0.57 & & 0.42 & 0.48 & 0.53 \\ 
$3z^2-r^2$ & 3.20 & 3.52 & 4.67 & 3.52 & 3.00 & & $3z^2-r^2$ & 0.69 & 0.42 & & 0.42 & 0.62 \\ 
$zx$ & 3.19 & 3.02 & 3.52 & 4.11 & 2.98 & & $zx$ & 0.57 & 0.48 & 0.42 & & 0.53 \\ 
$x^2-y^2$ & 3.49 & 2.98 & 3.00 & 2.98 & 3.78 & & $x^2-y^2$ & 0.32 & 0.53 & 0.62 & 0.53 & \\ 
\hline 
\hline 
FeTe &      &      & $U$   &      &          &  &     &     &      &  $J$   &      &    \\ [+1pt]
\hline \\ [-8pt] 
 & $xy$ & $yz$ & $3z^2-r^2$ & $zx$ & $x^2-y^2$   &   & & $xy$ & $yz$ & $3z^2-r^2$ & $zx$ & $x^2-y^2$  \\ 
\hline
$xy$ & 3.84 & 2.34 & 2.50 & 2.34 & 3.04 & & $xy$ & & 0.49 & 0.68 & 0.49 & 0.34 \\ 
$yz$ & 2.34 & 2.88 & 2.56 & 2.03 & 2.29 & & $yz$ & 0.49 & & 0.37 & 0.37 & 0.49 \\ 
$3z^2-r^2$ & 2.50 & 2.56 & 3.84 & 2.57 & 2.44 & & $3z^2-r^2$ & 0.68 & 0.37 & & 0.37 & 0.66 \\ 
$zx$ & 2.34 & 2.03 & 2.57 & 2.88 & 2.29 & & $zx$ & 0.49 & 0.37 & 0.37 & & 0.49 \\ 
$x^2-y^2$ & 3.04 & 2.29 & 2.44 & 2.29 & 3.59 & & $x^2-y^2$ & 0.34 & 0.49 & 0.66 & 0.49 & \\ 
\hline 
\hline 
\end{tabular} 
} 
\end{table*} 

\begin{table*}[htbp] 
\caption{Average of the diagonal terms of the screened Coulomb interaction, 
$\bar{U}$, and that of the bare Coulomb interaction $\bar{v}$. 
As a measure for the degree of screening, the value of 
$\bar{U}/\bar{v}$ is also shown. 
} 
\ 
\label{tab:uv} 
\begin{tabular}{l|ccc|ccc} 
\hline 
\hline 
\multicolumn{1}{c}{} & \multicolumn{3}{|c|}{$d$ model} & \multicolumn{3}{c}{$dp$/$dpp$ model}\\ \cline{2-7} \\ [-8pt]
 & $\bar{U}$ (eV)& $\bar{v}$ (eV)& $\bar{U}/\bar{v}$ & $\bar{U}$ (eV) & $\bar{v}$ (eV) & $\bar{U}/\bar{v}$  \\ 
\hline 
LaFePO & 2.47 & 14.15 & 0.174 & 4.13 & 18.96 & 0.218 \\
LaFeAsO & 2.53 & 14.85 & 0.171 & 4.23 & 19.46 & 0.217 \\ 
BaFe$_2$As$_2$ & 2.80 & 15.59 & 0.180 & 5.24 & 20.38 & 0.257 \\
LiFeAs & 3.15 & 15.82 & 0.199 & 5.94 & 20.35& 0.292 \\
FeSe & 4.24 & 17.53 & 0.242 & 7.21 & 21.37 & 0.337 \\
FeTe & 3.41 & 16.89 & 0.202 & 6.25 & 20.90& 0.299 \\
\hline 
\hline 
\end{tabular} 
\end{table*} 
Table~\ref{tab:uv} shows the average of the diagonal ($m=n$) terms of $U$. 
It is the smallest in the 1111 family. 
The value increases slightly by $\sim$0.3 eV in the BaFe$_2$As$_2$,  
and is substantially larger in the 111 and 11 families.  
The relative interaction strength $\bar{U}/\bar{t}$ in the $d$ model 
is nearly 8 for LaFePO, 9 for LaFeAsO, and 14 for FeSe. 
Therefore FeSe is substantially more strongly correlated.
Here, $\bar{U}$ is defined as the average of the diagonal intraorbital onsite interactions.
When we specify the largest transfer integral $t_{{\rm max}n}$ among pairs between an orbital $n$ at an iron site and any 3$d$ orbitals at its nearest-neighbor iron site, $\bar{t}$ is defined as the average of $t_{{\rm max}n}$ over the five orbitals; $\bar{t} \equiv \sum_{n=1}^5 t_{{\rm max}n}/5$.
For example, $\bar{t}$ for LaFeAsO is defined by 
(0.32+0.25+0.30+0.25+0.17)/5=0.26. 

Small (large) $U$ in the 1111 (11) family originates from the following two factors:
One is the MLWF basis for which $U$ is defined. 
The MLWF is more extended (localized) in the 1111 (11) family, 
thereby 
the bare interaction $\bar{v}$ becomes small (large). 
The other factor is the strength of the screening. 
As is shown in Table~\ref{tab:uv}, 
the ratio of the screened Coulomb interaction $\bar{U}$ to the bare interaction $\bar{v}$ is 
smaller in the 1111 family than in the 11 family. 
This can be understood as follows.  
Firstly, there is a larger number of bands near the Fe-$d$ band,  
such as occupied O-$p$ band and dense unoccupied states above the Fe-$d$ band.  
They contribute to screening in the 1111 systems. 
Secondly,  
the energy levels of the pnictogen-$p$ states in the 1111 systems are shallower 
than the Se-$p$ level in FeSe. It also enhances the screening.  
Hybridization between the $d$ and pnictogen/chalcogen orbitals would also 
affect the screening effect. 
As the hybridization becomes stronger, 
the transition matrix element would become larger,  
which makes screening more effective. 
 
There is little difference in $U$ between LaFePO and LaFeAsO,  
although $U$ is slightly larger in LaFeAsO.  
On the other hand, in the 11 system, FeSe is significantly  
more correlated than FeTe.  
The values of $\bar{U}$, $\bar{v}$, 
and $\bar{U}$/$\bar{v}$ 
are 24 \%, 4 \%, and 20 \% larger 
in FeSe than in FeTe, respectively. 
This means that the screening effect in $W_r$, 
not the size of the MLWFs, is curial for the difference.  
The chalcogen-$p$ states are shallower in FeTe than in FeSe. 
Transition energies between the $p$ states and unoccupied states are thus 
smaller, which results in stronger screening in FeTe. 
Also, the $p$ bands are entangled with the Fe-$d$ bands. 
This would change the transition matrix elements and leads to the
stronger screening. 

The $p$-$d$ hybridization delocalizes the Fe-3$d$ ${yz}$ and ${zx}$ Wannier orbitals in FeTe. 
In contrast to the other compounds, their spreads are the largest among the five orbitals. 
It is larger by 60 \% than that in FeSe, and close to that of LaFePO 
(see Table~\ref{tab:spread}). 
$U$ is accordingly smaller than other orbitals. 

We note that the value of $U$ in LaFeAsO is somewhat smaller than  
that in the previous works.~\cite{miyake08b,aichhorn09}  
We found that energy levels of the unoccupied states are sensitive to  
the number of basis functions taken into account in the FP-LMTO band calculations.  
Inclusion of more basis functions decreases $U$ slightly compared to the published data.  On top of that, we found that the presence of 
the localized La-4$f$ states in the 1111 family 
quantitatively affects the resulting $U$. 
In the quantitative aspect, in LDA, the La-$4f$ levels are expected to be too low, 
thus leading to artificially smaller $U$.  
In order to check this effect, we performed the cRPA calculation starting 
from the LDA+$U$ solution where the input-$U$ of 1 Ry is 
imposed on La-$4f$ states 
(this input-$U$ pushed up La-$4f$ level by $\sim$ 5 eV). 
The resulting effective onsite interaction $U$ for the Fe-3$d$ orbitals was found to increase roughly by 0.2 eV. 
We will come back to this point and 
examine it in more details later. 

The average of the exchange energies is 0.4 eV in the  
1111, 122, 111 families. In the 11 family,  
the value is slightly larger and is 0.5 eV.  
This trend is the same as the Coulomb interaction,  
though the family dependence is much weaker.  
As shown previously for the 3$d$ transition metals,~\cite{miyake08}  
screening effects are small for the exchange energy.  
The value is reduced only by $\sim$0.1 eV compared to the  
bare exchange interaction in all the compounds,  
in sharp contrast with the diagonal Coulomb interaction,  
where the reduction due to screening is a factor of 4-6.  
The family dependence 
primarily comes from the difference in the spatial extent of  
the MLWFs rather than the strength of screening.  
\begin{figure*}[htb]
\begin{center} 
\includegraphics[width=0.9\textwidth]{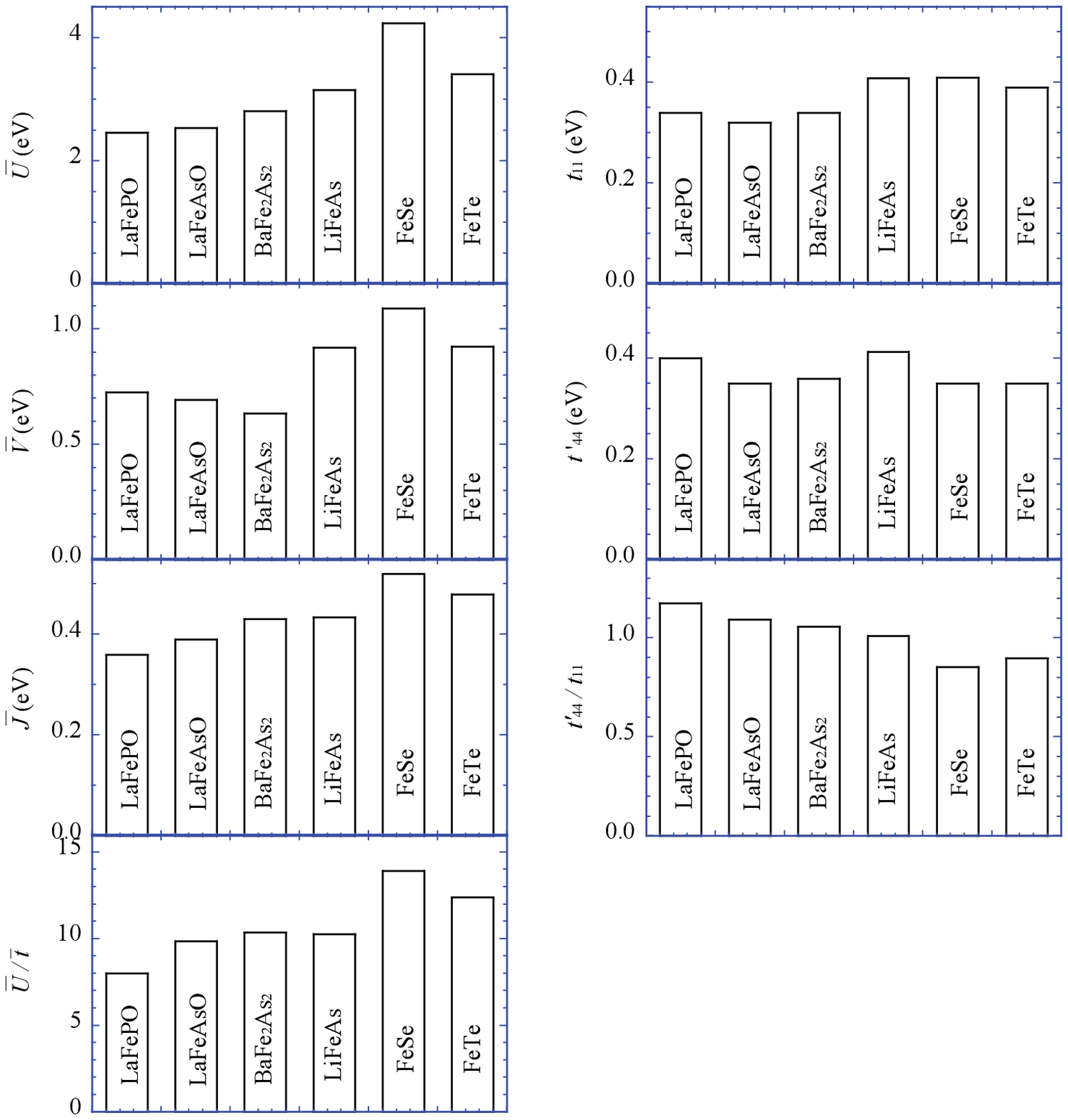}
\end{center} 
\caption{
Material dependence of parameters in $d$ model.  
The average of the onsite effective Coulomb interactions ($\bar{U}$),  
the average of the offsite effective Coulomb interactions between  
the neighboring Fe sites ($\bar{V}$),  
the average of the onsite effective exchange interactions ($\bar{J}$),  
the maximum value of the transfer integrals between  
the neighboring Fe sites [$t_{11} = t_{11}(1/2,-1/2,0)$]
 and between the next-nearest neighbor [$t'_{44} = t_{44}(1,0,0)$],  
$\bar{U}/\bar{t}$, and $t'_{44}/t_{11}$ 
are compared.  
The subscripts of $t_{11}$ and $t'_{44}$ are orbital indices; 1 for $xy$ and 4 for $zx$.
$\bar{t}$ is the orbital average of the largest nearest $d$-$d$ transfer integrals. 
}
\label{fig:summary} 
\end{figure*} 

Interaction between the nearest neighbor Fe sites, $V$, 
is shown in Table \ref{tab:uu_d}. 
The value of $V$ is 0.6-1.1 eV, and  
the orbital dependence of $V$ is weak 
in contrast to the onsite $U$. 
The crucial factor for $V$ is 
not the spatial extent of the MLWFs but the distance between the sites.   
Longer-range interactions are found to decay as $1/\alpha r$, where $r$ is the distance between  
the Wannier centers and 
$\alpha$=0.5-0.6 (eV$\cdot\AA)^{-1}$ in the 1111 and 122 families, and 
$\alpha$=0.3.-0.4 (eV$\cdot\AA)^{-1}$ in the 11 and 111 families.

The presented results for the $d$ model 
are summarized in Fig.~\ref{fig:summary}.
From this plot, the systematic change of the screened interaction and the geometrical frustration 
measured from the amplitude of the next-nearest-neighbor to the nearest-neighbor transfers are 
more visible with the evolution from LaFePO, LaFeAsO, BaFe$_{2}$As$_2$, LiFeAs to 
FeSe and FeTe.
We again emphasize that the family dependence of $\bar{U}/\bar{t}$ comes from $\bar{U}$ not from $\bar{t}$. 

\begin{table*}[htb] 
\caption{Effective Coulomb interaction between the neighboring  
sites, $V$, in the $d$ model (in eV).  
} 
\ 
\label{tab:uu_d} 
{\scriptsize 
\begin{tabular}{ccccccccccccccc} 
\hline \hline \\ [-8pt]
LaFePO & $xy$ & $yz$ & $3z^2-r^2$ & $zx$ & $x^2-y^2$  & & LaFeAsO & $xy$ & $yz$ & $3z^2-r^2$ & $zx$ & $x^2-y^2$  \\ 
\hline 
$xy$ & 0.77 & 0.74 & 0.74 & 0.74 & 0.74 & & $xy$ & 0.73 & 0.70 & 0.69 & 0.70 & 0.71 \\ 
$yz$ & 0.74  & 0.72 & 0.72 & 0.73 & 0.71 & & $yz$ & 0.70 & 0.68 & 0.68 & 0.69 & 0.69 \\ 
$3z^2-r^2$ & 0.74  & 0.72 & 0.71 & 0.72 & 0.71 & & $3z^2-r^2$ & 0.69 & 0.68 & 0.67 & 0.68 & 0.68 \\ 
$zx$ & 0.74  & 0.73 & 0.72 & 0.72 & 0.71 & & $zx$ & 0.70 & 0.69 & 0.68 & 0.68 & 0.69 \\ 
$x^2-y^2$ & 0.74 & 0.71 & 0.71 & 0.71 & 0.71 & & $x^2-y^2$ & 0.71 & 0.69 & 0.68 & 0.69 & 0.71 \\ 
\hline 
\hline 
BaFe$_2$As$_2$ & $xy$ & $yz$ & $3z^2-r^2$ & $zx$ & $x^2-y^2$ & & LiFeAs & $xy$ & $yz$ & $3z^2-r^2$ & $zx$ & $x^2-y^2$  \\ 
\hline 
$xy$ & 0.66 & 0.63 & 0.64 & 0.63 & 0.66 & & $xy$ & 0.94 & 0.92 & 0.91 & 0.92 & 0.93 \\ 
$yz$ & 0.63 & 0.61 & 0.62 & 0.63 & 0.63 & & $yz$ & 0.92 & 0.91 & 0.90 & 0.92 & 0.92 \\ 
$3z^2-r^2$ & 0.64 & 0.62 & 0.62 & 0.62 & 0.63 & & $3z^2-r^2$ & 0.91 & 0.90 & 0.89 & 0.90 & 0.91 \\ 
$zx$ & 0.63 & 0.63 & 0.62 & 0.61 & 0.63 & & $zx$ & 0.92 & 0.92 & 0.90 & 0.91 & 0.92 \\ 
$x^2-y^2$ & 0.66 & 0.63 & 0.63 & 0.63 & 0.66 & & $x^2-y^2$ & 0.93 & 0.92 & 0.91 & 0.92 & 0.95 \\ 
\hline 
\hline 
FeSe & $xy$ & $yz$ & $3z^2-r^2$ & $zx$ & $x^2-y^2$ & & FeTe & $xy$ & $yz$ & $3z^2-r^2$ & $zx$ & $x^2-y^2$  \\ 
\hline 
$xy$ & 1.11 & 1.09 & 1.08 & 1.09 & 1.11 & &  $xy$ & 0.94 & 0.92 & 0.92 & 0.92 & 0.94 \\ 
$yz$ & 1.09 & 1.08 & 1.07 & 1.10 & 1.09 & & $yz$ & 0.92 & 0.91 & 0.91 & 0.92 & 0.91 \\ 
$3z^2-r^2$ & 1.08 & 1.07 & 1.06 & 1.07 & 1.08 & & $3z^2-r^2$ & 0.92 & 0.91 & 0.91 & 0.91 & 0.91 \\ 
$zx$ & 1.09 & 1.10 & 1.07 & 1.08 & 1.09 & & $zx$ & 0.92 & 0.92 & 0.91 & 0.91 & 0.91 \\ 
$x^2-y^2$ & 1.11 & 1.09 & 1.08 & 1.09 & 1.11 & & $x^2-y^2$ & 0.94 & 0.91 & 0.91 & 0.91 & 0.94 \\ 
\hline 
\hline 
\end{tabular} 
} 
\end{table*} 

\subsection{Transfer integrals of $dp/dpp$ model}
The effective parameters are very different in the $dpp$ (in LaFePO and LaFeAsO) or $dp$ (in other materials) 
model compared to those in the $d$ model. 
The change in the spread and shape of the MLWFs 
(compare Figs.~\ref{fig:mlwf_d} and \ref{fig:mlwf_dp}) 
results in the change in the transfer integrals. 
The largest transfer integral between 
the nearest-neighbor $d$ orbitals, $t_{\rm max}$, is 
the $d_{xy}$-$d_{xy}$ hopping. 
Its value is comparable or even larger in the $dpp$/$dp$ model;  
0.316 eV, 0.403 eV, 0.367 eV for LaFeAsO, FeSe and FeTe, respectively, 
against 0.315 eV, 0.410 eV, 0.392 eV in the $d$ model.
Other nearest $d$-$d$ transfer integrals 
alter significantly, because the change in 
the MLWFs is significant. Most of them are 
smaller in magnitude in the $dpp$/$dp$ model. 
However, the $d_{x^2-y^2}$-$d_{x^2-y^2}$ transfer integral gets larger. 
The value is 
0.243 eV, 0.272 eV, 0.308 eV for LaFeAsO, FeSe, FeTe, respectively in the $dpp$/$dp$ model, 
while it is 0.152 eV, 0.057 eV, 0.085 eV in the $d$ model. 
 The next-nearest-neighbor $d$-$d$ transfer integrals are small
 and even the
 largest value is $\sim$0.1 eV in all the materials, to be compared with 0.3-0.4 eV in the $d$ model, clearly indicating that the next-neighbor transfers 
 in the $d$ model is mediated by the pnictogen-/chalcogen-$p$ orbitals. 
 In fact, in the $dp/dpp$ model, the transfer integrals 
 between the Fe-$d$ and pnictogen-/chalcogen-$p$ orbitals are 
 larger than the $d$-$d$ transfer integrals;  
 the largest one is the $d_{zx}$-$p_{x}$ transfer integral, 
 the value of which is 
 0.734 eV, 0.895 eV, and 0.711 eV for LaFeAsO, 
 FeSe and FeTe, respectively. 
 It should be noted here that,  
 in comparison of the three compounds, the value is the largest in FeSe. 
This is unexpected; 
the spread of the $d_{zx}$ orbital is the largest 
and the distance between Fe and pnictogen/chalcogen atoms is the shortest in LaFeAsO. 
Therefore, naively we expect that the the transfer integral would be the largest in LaFeAsO, 
which turned out not to be the case. 
Concerning the $p$-$p$ transfer integrals between the nearest pnictogen/chalcogen sites,
the largest value is for the $p_z$-$p_z$ transfer, which is 0.306 eV, 0.301 eV, 0.319 eV
in LaFeAsO, FeSe, and FeTe, respectively. The $p_x$-$p_x$ ($p_y$-$p_y$) hopping is 
0.255 eV, 0.203 eV and 0.164 eV, and the $p_x$-$p_z$ ($p_y$-$p_z$) hopping is 
0.279 eV, 0.272 eV and 0.225 eV in each material.

\subsection{Screened Coulomb interaction of $dp/dpp$ model}
Table~\ref{tab:uj_dp} presents the $U$ and $J$ matrices
in the $dp/dpp$ model.
The value of $U$ is large compared to the $d$ model for the  
following two reasons.\cite{miyake08,aichhorn09} 
Firstly, there are more states kept in the model, so that 
more screening processes are taken away from $W_r$.  
Secondly, the MLWFs are more localized, since they are optimized using more states. 
In addition, the MLWFs in the $dpp$/$dp$ model is less orbital dependent. 
Consequently, the orbital dependence in $U$ 
is much weaker than the $d$ model.  
The exchange value is somewhat 
larger than that in the $d$ model by $\sim$0.2 eV. 

Examining the family dependence, 
we find that 
$U$ is substantially larger in the 11 family than that in the 1111 family: 
The value of $\bar{U}/\bar{t}$ ($\bar{U}/W$) is 20 (0.48), 20 (0.58), 30 (0.91), 26 (0.77) in LaFePO, LaFeAsO, FeSe, FeTe, respectively. 
Here, $\bar{U}$ is the average of the diagonal onsite interactions between the $d$ orbitals, 
and $W$ is the width of the $d$ band. 
We take the definition of $\bar{t}$ as the same as the case of the $d$ model;
$\bar{t}$ is defined by the largest transfers between the nearest-neighbor iron sites averaged over the 3$d$ orbitals.  
As is shown in Table~\ref{tab:uv}, 
the bare Coulomb interaction $\bar{v}$ differs 
only by 10 \% in the two families, 
while ${\bar U}/{\bar v}$ of the 1111 family is smaller by a factor more than 25 \% than that of the 1111 family. 
Therefore, the family dependence of $U$ prominent in Table~\ref{tab:uv} is determined primarily 
by the difference in the screening while the spatial extent of MLWFs plays an only supplementary role. 
\begin{table*}[htb] 
\caption{Effective Coulomb ($U$) / exchange ($J$) interaction 
between two electrons at 3$d$ orbitals on the same iron site
in the $dpp$ (for LaFePO and LaFeAsO) or $dp$ (in other compounds) model (in eV).  
} 
\ 
\label{tab:uj_dp} 
{\scriptsize 
\begin{tabular}{ccccccccccccccc} 
\hline \hline \\ [-8pt]
LaFePO &      &      & $U$   &      &          &  &     &     &      &  $J$   &      &    \\ [+1pt]
\hline \\ [-8pt] 
 & $xy$ & $yz$ & $3z^2-r^2$ & $zx$ & $x^2-y^2$   &   & & $xy$ & $yz$ & $3z^2-r^2$ & $zx$ & $x^2-y^2$  \\ 
\hline
$xy$ & 4.72 & 3.09 & 2.97 & 3.09 & 3.47 & & $xy$ & & 0.63 & 0.72 & 0.63 & 0.35 \\ 
$yz$ & 3.09 & 3.99 & 3.25 & 2.85 & 2.81 & & $yz$ & 0.63 & & 0.43 & 0.54 & 0.55 \\ 
$3z^2-r^2$ & 2.97 & 3.25 & 4.25 & 3.25 & 2.72 & & $3z^2-r^2$ & 0.72 & 0.43 & & 0.43 & 0.62 \\ 
$zx$ & 3.09 & 2.85 & 3.25 & 3.99 & 2.81 & &  $zx$ & 0.63 & 0.54 & 0.43 & & 0.55 \\ 
$x^2-y^2$ & 3.47 & 2.81 & 2.72 & 2.81 & 3.71 & & $x^2-y^2$ & 0.35 & 0.55 & 0.62 & 0.55 & \\ 
\hline 
\hline 
LaFeAsO &  &  & $U$   &      &          &  &     &     &      &  $J$   &      &    \\ [+1pt]
\hline \\ [-8pt] 
 & $xy$ & $yz$ & $3z^2-r^2$ & $zx$ & $x^2-y^2$   &   & & $xy$ & $yz$ & $3z^2-r^2$ & $zx$ & $x^2-y^2$  \\ 
\hline
$xy$ & 4.66 & 3.09 & 2.99 & 3.09 & 3.57 & & $xy$ & & 0.63 & 0.74 & 0.63 & 0.37  \\ 
$yz$ & 3.09 & 4.08 & 3.31 & 2.90 & 2.91 & & $yz$ & 0.63 & & 0.45 & 0.56 & 0.59 \\ 
$3z^2-r^2$ & 2.99 & 3.31 & 4.33 & 3.31 & 2.81 & & $3z^2-r^2$ & 0.74 & 0.45 & & 0.45 & 0.67 \\ 
$zx$ & 3.09 & 2.90 & 3.31 & 4.08 & 2.91 &  & $zx$ & 0.63 & 0.56 & 0.45 & & 0.59 \\ 
$x^2-y^2$ & 3.57 & 2.91 & 2.81 & 2.91 & 3.98 & & $x^2-y^2$ & 0.37 & 0.59 & 0.67 & 0.59 & \\  
\hline 
\hline 
BaFe$_2$As$_2$ &      &      & $U$   &      &     &  &     &     &      &  $J$   &      &    \\ [+1pt]
\hline \\ [-8pt] 
 & $xy$ & $yz$ & $3z^2-r^2$ & $zx$ & $x^2-y^2$   &   & & $xy$ & $yz$ & $3z^2-r^2$ & $zx$ & $x^2-y^2$  \\ 
\hline
$xy$ & 5.40 & 3.95 & 3.84 & 3.95 & 4.40 & & $xy$ & & 0.68 & 0.78 & 0.68 & 0.39 \\ 
$yz$ & 3.95 & 5.19 & 4.33 & 3.86 & 3.81 & & $yz$ & 0.68 & & 0.49 & 0.64 & 0.66 \\ 
$3z^2-r^2$ & 3.84 & 4.33 & 5.45 & 4.33 & 3.71 & & $3z^2-r^2$ & 0.78 & 0.49 & & 0.49 & 0.75 \\ 
$zx$ & 3.95 & 3.86 & 4.33 & 5.19 & 3.81 & & $zx$ & 0.68 & 0.64 & 0.49 & & 0.66 \\ 
$x^2-y^2$ & 4.40 & 3.81 & 3.71 & 3.81 & 4.97 & & $x^2-y^2$ & 0.39 & 0.66 & 0.75 & 0.66 & \\ 
\hline 
\hline 
LiFeAs &      &      & $U$   &      &     &  &     &     &      &  $J$   &      &    \\ [+1pt]
\hline \\ [-8pt] 
 & $xy$ & $yz$ & $3z^2-r^2$ & $zx$ & $x^2-y^2$   &   & & $xy$ & $yz$ & $3z^2-r^2$ & $zx$ & $x^2-y^2$  \\ 
\hline
$xy$ & 5.98 & 4.60 & 4.49 & 4.60 & 5.14 & & $xy$ & & 0.67 & 0.77 & 0.67 & 0.39 \\ 
$yz$ & 4.60 & 5.89 & 5.01 & 4.55 & 4.58 & & $yz$ & 0.67 & & 0.48 & 0.63 & 0.67 \\ 
$3z^2-r^2$ & 4.49 & 5.01 & 6.08 & 5.01 & 4.46 & & $3z^2-r^2$ & 0.77 & 0.48 & & 0.48 & 0.77 \\ 
$zx$ & 4.60 & 4.55 & 5.01 & 5.89 & 4.58 & & $zx$ & 0.67 & 0.63 & 0.48 & & 0.67 \\ 
$x^2-y^2$ & 5.14 & 4.58 & 4.46 & 4.58 & 5.87 & & $x^2-y^2$ & 0.39 & 0.67 & 0.77 & 0.67 & \\ 
\hline 
\hline 
FeSe &      &      & $U$   &      &     &  &     &     &      &  $J$   &      &    \\ [+1pt]
\hline \\ [-8pt] 
 & $xy$ & $yz$ & $3z^2-r^2$ & $zx$ & $x^2-y^2$   &   & & $xy$ & $yz$ & $3z^2-r^2$ & $zx$ & $x^2-y^2$  \\ 
\hline
$xy$ & 7.21 & 5.76 & 5.56 & 5.76 & 6.30 & & $xy$ & & 0.74 & 0.83 & 0.74 & 0.42 \\ 
$yz$ & 5.76 & 7.25 & 6.18 & 5.75 & 5.73 & & $yz$ & 0.74 & & 0.53 & 0.71 & 0.74 \\ 
$3z^2-r^2$ & 5.56 & 6.18 & 7.23 & 6.18 & 5.52 & & $3z^2-r^2$ & 0.83 & 0.53 & & 0.53 & 0.83 \\ 
$zx$ & 5.76 & 5.75 & 6.18 & 7.25 & 5.73 & & $zx$ & 0.74 & 0.71 & 0.53 & & 0.74 \\ 
$x^2-y^2$ & 6.30 & 5.73 & 5.52 & 5.73 & 7.09 & & $x^2-y^2$ & 0.42 & 0.74 & 0.83 & 0.74 & \\ 
\hline 
\hline 
FeTe &      &      & $U$   &      &     &  &     &     &      &  $J$   &      &    \\ [+1pt]
\hline \\ [-8pt] 
 & $xy$ & $yz$ & $3z^2-r^2$ & $zx$ & $x^2-y^2$   &   & & $xy$ & $yz$ & $3z^2-r^2$ & $zx$ & $x^2-y^2$  \\ 
\hline
$xy$ & 6.09 & 4.81 & 4.60 & 4.81 & 5.42 & & $xy$ & & 0.69 & 0.78 & 0.69 & 0.41 \\ 
$yz$ & 4.81 & 6.29 & 5.23 & 4.85 & 4.91 & & $yz$ & 0.69 & & 0.50 & 0.69 & 0.72 \\ 
$3z^2-r^2$ & 4.60 & 5.23 & 6.18 & 5.23 & 4.69 & & $3z^2-r^2$ & 0.78 & 0.50 & & 0.50 & 0.81 \\ 
$zx$ & 4.81 & 4.85 & 5.23 & 6.29 & 4.91 & & $zx$ & 0.69 & 0.69 & 0.50 & & 0.72 \\ 
$x^2-y^2$ & 5.42 & 4.91 & 4.69 & 4.91 & 6.37 & & $x^2-y^2$ & 0.41 & 0.72 & 0.81 & 0.72 & \\ 
\hline 
\hline 
\end{tabular} 
} 
\end{table*} 

The effective onsite Coulomb interaction 
of the pnictogen-/chalcogen-$p$ orbitals,  
$U_p$, are also 
large (not shown in the table).  
It is the smallest in LaFePO, 
still the average of the diagonal terms is 2.5 eV.  
The value is the largest for FeSe at 4.7 eV.  
Interaction between the Fe-$d$ and neighboring-pnictogen-$p$ orbitals,  
$U_{pd}$, is not negligible as well,  
ranging from 1.2 eV (LaFeAsO) to 1.7 eV (FeSe).

\subsection{Comparison between FP-LMTO and pseudopotential calculations}
 For the critical check of the reliability 
 and convergence of the derived parameters, we compared   
 the FP-LMTO results with results obtained from {\em ab initio} 
 pseudopotential calculations 
 with the plane-wave basis.  
 Comparisons were made for onsite $U$ and 
 $J$ parameters of LaFeAsO and FeSe. 
 In LaFeAsO, the pseudopotential results (top two left 5$\times$5 matrices 
 in Table~\ref{tab:nakamura:U}) give  
 somewhat larger values by 0.1-0.2 eV than the FP-LMTO values listed in  
  Table~\ref{tab:uj_d}. 
 This is because the present pseudopotential calculation 
 does not include La-$f$ states and therefore the screenings  
 from the La-$f$ states were completely neglected in the RPA calculations. 
 On the other hand, the FP-LMTO results include the 
 La-$f$-screening effects and thus  
 give reasonably smaller values than the pseudopotential ones.  
 It should be noted here that, in general, LDA tends to underestimate  
the energy difference between 
 the level of the localized $f$ state and the Fermi level.  
 In the present calculation, this level is located around  
 3 eV above the Fermi level [see Fig.~\ref{fig:band}~(a)],  
 which may be too low. 
 In order to analyze the effect of the La-$f$ level on the  
 derived parameters, we performed constrained RPA 
 calculations in LDA+$U$ formalism with the FP-LMTO implementation,   
 where we employed 1 Ry for 
input-$U$ on the La-$f$ orbitals, 
which pushed the $f$ level up by nearly 5 eV 
from the original position. 
 The resultant $U$ and $J$ parameters are shown in 
 the top two right 5$\times$5 matrices in 
 Table~\ref{tab:nakamura:U}, from which   
 we see that, with this input-$U$, LDA+$U$
 gives values very similar to
 the pseudopotential results and consistently suggests this possible small correction ($\sim 0.2$ eV). 
 
 In contrast, in FeSe, there exists no ambiguity due to  
 the La-$f$ state;    
 the pseudopotential results should agree 
 with the original FP-LMTO results without resorting to LDA+$U$. 
 Comparisons are shown in the bottom two part  
 in Table~\ref{tab:nakamura:U}. 
 We see an excellent agreement between the two results, 
 which confirms that the constrained RPA results based on  
 the MLWFs   
 are neither affected by details of treatments of core electrons, 
 nor by the basis choices ({\it i.e.} either plane waves or FP-LMTO),  
if the cutoff radius in the pseudopotential 
 is taken small enough, and reasonably large number of unoccupied 
 states are included in the cRPA calculation. 

We here note on the convergence of the cRPA. 
In the cRPA, the number of bands participating in the screening 
should be taken sufficiently large in the part away from the Fermi level,
so that the polarization calculation converges.  
In the present pseudopotential calculations, 
the total number of conduction bands, $N_{band}$, 
is taken up to 130 together with the number of the valence bands, 
 34 for LaFeAsO and 18 for FeSe.  
With this choice of the numbers, we infer that the underestimate error of the polarization may lead to 
an overestimate of the diagonal part of the screened interaction with the amount at most 0.1 eV.
(For example, for $U_{xy}$ of LaFeAsO, 
the value is 3.43 eV for $N_{band} = 50$, 
 3.21 eV for  $N_{band} = 70$, and 3.14 eV for $N_{band} = 130$.)
If one wishes to reach better accuracy, 
one needs to take larger number of bands, 
while for the present purpose with the accuracy of the order of 0.1 eV, the present choice may be sufficient.

\begin{table*}[htb] 
\caption{Effective onsite Coulomb/exchange interactions 
in the $d$ model obtained with {\em ab initio} 
pseudopotential (PP) calculations, compared with 
the results by the FP-LMTO. The FP-LMTO calculation for 
LaFeAsO is done using the LDA+$U$ band structure 
with input-$U$ of 1 Ry for the La-$4f$ orbitals, 
leading to a shift of the La-4$f$ level by 
$\sim$ 5 eV from the original position.} 
\label{tab:nakamura:U} 
\  
\  
{\scriptsize 
\begin{tabular}{ccccccccccccccc} 
\hline \hline \\ [-8pt] 
LaFeAsO &      &      &     PP     &      &   &       &      &      &   &  FP-LMTO    &         \\ [+1pt] 
\hline \\ [-8pt] 
 $U$ & $xy$ & $yz$ & $3z^2-r^2$ & $zx$ & $x^2-y^2$   &   & $U$ & $xy$ & $yz$ & $3z^2-r^2$ & $zx$ & $x^2-y^2$  \\ 
\hline \\ [-8pt]  
 $xy$     & 3.14  & 1.97 & 1.99 & 1.97 & 2.03  &   &  
 $xy$     & 3.16  & 1.93 & 1.91 & 1.93 & 2.03  \\  
 $yz$     & 1.97  & 2.67 & 2.13 & 1.83 & 1.70  &   &  
 $yz$     & 1.93  & 2.63 & 2.14 & 1.77 & 1.65  \\ 
 $3z^2-r^2$& 1.99  & 2.13 & 3.09 & 2.13 & 1.72  &   &  
 $3z^2-r^2$& 1.91  & 2.14 & 3.05 & 2.14 & 1.64  \\ 
 $zx$      & 1.97  & 1.83 & 2.13 & 2.67 & 1.70  &   &  
 $zx$      & 1.93  & 1.77 & 2.14 & 2.63 & 1.65  \\ 
 $x^2-y^2$ & 2.03  & 1.70 & 1.72 & 1.70 & 2.11  &   &  
 $x^2-y^2$ & 2.03  & 1.65 & 1.64 & 1.65 & 2.05  \\ 
 \hline \\ [-8pt] 
 $J$ & $xy$ & $yz$ & $3z^2-r^2$ & $zx$ & $x^2-y^2$   &   &  $J$ & $xy$ & $yz$ & $3z^2-r^2$ & $zx$ & $x^2-y^2$   \\ 
 \hline \\ [-8pt]  
 $xy$      &       & 0.46 & 0.56 & 0.46 & 0.26  & &  
 $xy$      &       & 0.48 & 0.58 & 0.48 & 0.24  \\ 
 $yz$      & 0.46  &      & 0.35 & 0.39 & 0.37  & &  
 $yz$      & 0.48  &      & 0.35 & 0.39 & 0.37  \\ 
 $3z^2-r^2$     & 0.56  & 0.35 &      & 0.35 & 0.43  & &  
 $3z^2-r^2$     & 0.58  & 0.35 &      & 0.35 & 0.44  \\ 
 $zx$      & 0.46  & 0.39 & 0.35 &      & 0.37  & &  
 $zx$      & 0.48  & 0.39 & 0.35 &      & 0.37  \\ 
 $x^2-y^2$ & 0.26  & 0.37 & 0.43 & 0.37 &       & &  
 $x^2-y^2$ & 0.24  & 0.37 & 0.44 & 0.37 &       \\ 
\hline \hline \\ [-8pt] 
 FeSe   &      &      &     PP     &      &     &   &       &      &      &    FP-LMTO    &      &   \\ [+1pt]  
\hline \\ [-8pt]  
  $U$ & $xy$ & $yz$ & $3z^2-r^2$ & $zx$ & $x^2-y^2$  & &  $U$ & $xy$ & $yz$ & $3z^2-r^2$ & $zx$ & $x^2-y^2$  \\ 
 \hline \\ [-8pt]  
 $xy$      & 4.35 & 3.10 & 3.15 & 3.10 & 3.30   & & 
 $xy$      & 4.51 & 3.19 & 3.20 & 3.19 & 3.49   \\ 
 $yz$      & 3.10 & 3.97 & 3.36 & 2.95 & 2.91   & &  
 $yz$      & 3.19 & 4.11 & 3.52 & 3.02 & 2.98   \\ 
 $3z^2-r^2$     & 3.15 & 3.36 & 4.50 & 3.36 & 2.97   & &  
 $3z^2-r^2$     & 3.20 & 3.52 & 4.67 & 3.52 & 3.00   \\ 
 $zx$      & 3.10 & 2.95 & 3.36 & 3.97 & 2.91   & & 
 $zx$      & 3.19 & 3.02 & 3.52 & 4.11 & 2.98   \\ 
 $x^2-y^2$ & 3.30 & 2.91 & 2.97 & 2.91 & 3.68   & & 
 $x^2-y^2$ & 3.49 & 2.98 & 3.00 & 2.98 & 3.78   \\ 
 \hline \\ [-8pt]  
  $J$ & $xy$ & $yz$ & $3z^2-r^2$ & $zx$ & $x^2-y^2$  & & $J$ & $xy$ & $yz$ & $3z^2-r^2$ & $zx$ & $x^2-y^2$  \\ 
 \hline \\ [-8pt]  
 $xy$      &      & 0.54 & 0.65 & 0.54 & 0.32   & &  
 $xy$      &      & 0.57 & 0.69 & 0.57 & 0.32   \\ 
 $yz$      & 0.54 &      & 0.41 & 0.47 & 0.50   & & 
 $yz$      & 0.57 &      & 0.42 & 0.48 & 0.53   \\ 
 $3z^2-r^2$     & 0.65 & 0.41 &      & 0.41 & 0.58   & &
 $3z^2-r^2$     & 0.69 & 0.42 &      & 0.42 & 0.62   \\ 
 $zx$      & 0.54 & 0.47 & 0.41 &      & 0.50   & & 
 $zx$      & 0.57 & 0.48 & 0.42 &      & 0.53   \\ 
 $x^2-y^2$ & 0.32 & 0.50 & 0.58 & 0.50 &        & & 
 $x^2-y^2$ & 0.32 & 0.53 & 0.62 & 0.53 &        \\ 
 \hline \hline  
\end{tabular}
} 
\end{table*} 

Finally, we briefly comment an effect of replacement of LDA 
 by GGA on the derived parameters.  
We found that such a replacement hardly affects the resultant 
parameters; GGA gives slightly larger  
values than those obtained with LDA. 
The difference is less than 0.05 eV 
at the maximum and thus   
 the choice of LDA or GGA would not be essential  
 in the derivation of the  
 model parameters.  

\section{Summary and Discussion}\label{Summary} 
 
In this paper, we have derived effective low-energy models of iron-based superconductors, 
by applying the three-stage scheme including the downfolding to eliminate the higher-energy excitation 
channels and to retain the low-energy (target bands) degrees of freedom near the Fermi level.  
The models derived in this paper are those for LaFePO, and LaFeAsO belonging to the 1111 family, 
BaFe$_2$As$_2$ (122) and LiFeAs (111) as well as FeSe and FeTe belonging to the 11 family. 
For each compound, two different types of low-energy models are presented: 
One, called the $d$ model, is constructed from the bands which contain mainly Fe-$3d$ bands only.  
In this case, the model contains 10 orbitals per unit cell containing two iron sites.  
In the other model called the $dp$ model, 
As-4$p$/P-3$p$/Se-4$p$/Te-5$p$ orbitals are additionally included in the low-energy models. 
In case of LaFePO and LaFeAsO, O-$2p$ orbitals are included as well, which constitute the $dpp$ models.   
The downfolding procedure starts from the LDA calculation results of the global band structures.  
The MLWFs are constructed for the target low-energy bands.  
The screened Coulomb interaction between two electrons in the target bands are derived by the constrained RPA, 
where the screening effects arising from the eliminated bands are included.  
The effective low-energy models are derived from the two independent LDA calculations, one based on 
the FP-LMTO and the other based on the pseudopotential with the plane wave basis. 
These two methods give very good agreements and the resulting model parameters do not depend on the choice of 
the basis function in the LDA calculations, 
which assures the reliability of 
 the downfolding scheme presented here.  
 We also checked that it neither depends on the choice of LDA or GGA.  

In addition, in the constrained calculations, we have applied 
a recently developed technique to the disentanglement of bands~\cite{Miyake} 
in the procedure of the downfolding to disconnect the target band from other entangled band structure. 
This is particularly helpful for  
FeTe, where Fe-3$d$ bands are entangled with the Te-$p$ bands. 

The band structures of the six compounds 
share global similarity. 
All of them at the Fermi level show small electron pockets around the M point and small hole pockets around the $\Gamma$ point. \
However, there exist nonnegligible differences and dependence of the derived model parameters on the compounds and families. 
We start from the overall family dependence that does not depend on the choice of the $d$ or $dp/dpp$ models. 
One important origin of the family dependence of the model parameters is 
the variation of the distance $h$ between the pnictogen or chalcogen atoms and the Fe layer, as listed in Table \ref{tab:geometry}. 
In the ascending order, the distance becomes longer from the 1111, 111, 122 to 11 families. 
The distance is the shortest 
 for the LaFePO and sequentially increases 
 in the direction from LaFeAsO, LiFeAs, 
 BaFe$_2$As$_2$ to FeSe and FeTe. 
 Then, in this order, the compounds progressively lose covalent character 
  of the chemical bonding between Fe and the chalcogen/pnictogen elements and gain stronger ionic character. 
In this view, FeTe is expected to have the largest ionic character. 
On top of that, electronegativity 
 is also important for analyses of the chemical bonding.
The value of electronegativity for each element is 1.8 for Fe, 2.1 for P, 2.0 for As, 2.4 for Se, and 2.1 for Te.
So, in this view, the bonding of FeSe has the largest ionic character.
By contrast to FeSe and FeTe, the 1111 family has relatively strong hybridization between Fe-3$d$ and pnictogen-$p$ orbitals. 
Indeed, the difference of the distance $h$, as well as the electronegativity, explains many of the family dependence as revealed below 
 in a unified fashion.  
 
The 11 family, in particular, FeTe has more entangled band structure of 
the Fe-$3d$ and chalcogen-$p$ bands, leading to the smearing of the  pseudogap structure within the 3$d$ bands above the Fermi level in contrast to the prominent pseudogap (located at $\sim 0.5$ eV higher than the Fermi level) in the 1111 family. 
In terms of the $d$ model, the origin of the pseudogap 
structure can be understood as follows. Let us first see
why the ${3z^2-r^2}$ and ${xy}$ bands have a gap around the Fermi level 
commonly for the four families (see Fig. \ref{fig:pdos}). 
The energy scale (band width) of the ${xy}$ band is the largest among 
the 3$d$ bands, since the ${xy}$ orbital points toward neighboring 
Fe sites. 
 As is well appreciated, any two-dimensional 
single-orbital tight-binding
models do not make 
gaps in the DOS unless the translational symmetry is broken.
On the other hand, as mentioned in the previous section, 
 all the families of the iron compounds have 
a high translational symmetry (for the appropriate 
choice of the sublattice dependent gauge). 
Therefore, the gap in the ${xy}$ band is not the single-band origin of the sole $xy$ orbital.  
Instead, the origin is explicable only
 with the hybridization between ${xy}$ and
other 3$d$ orbitals.
As we can see in Tables \ref{t_1111}-\ref{t_11}, 
the nearest-neighbor transfer integrals 
[$t_{mn}(\frac{1}{2},\frac{1}{2},0)$] between $m=xy$ and $n=yz$, $zx$,  
$3z^2-r^2$ are appreciable; they are all $\sim$ 300 meV. 
 It should be noted here that, 
 even the ${3z^2-r^2}$ orbital perpendicular to the iron layer has rather  
 large hybridization; $t_{xy,3z^2-r^2}(\frac{1}{2},\frac{1}{2},0)$ 
 $\sim$ 300 meV.
 Through this hybridization, 
 the ${xy}$ and ${3z^2-r^2}$ bands tend to make a clear gap 
 around the Fermi level. 

For the ${x^2-y^2}$ orbital, the nearest 
$t_{xy,x^2-y^2}(\frac{1}{2},\frac{1}{2},0)$ 
is always small, but the next-nearest $t_{x^2-y^2,3z^2-r^2}(1,0,0)$ 
can be large for 
systems with smaller $h$. For example, $t_{x^2-y^2,3z^2-r^2}(1,0,0)$ 
is just $-$23 meV for FeTe but $-$234 meV for LaFePO.
This is because the Wannier spread 
of the ${x^2-y^2}$ orbital becomes large for smaller $h$ (see Table \ref{tab:spread}). 
The pseudogap structure in the ${x^2-y^2}$ band thus appears 
for small-$h$ system.
We note that the two-peak structure is observed in 
the ${x^2-y^2}$ DOS even without interorbital hybridization, 
as shown in the right panel of Fig.\ref{fig:pdos}. 
However, the energy level of the lower peak is shifted up in that case, 
thus the pseudogap is not formed at the Fermi level.
Also, the energy separation between the two peaks is too small.
Analysis using the $dpp/dp$ model even more clearly reveals that the psudogap indeed originates from the hybridization gap, indicating the significance of
the interorbital hybridization : Without the hybridization between the $x^2-y^2$ and the anion $p$ orbitals, which is included in the $x^2-y^2$ Wannier in the $d$ model,  the upper peak and dip structure itself disappears. 

A similar argument can be applied to ${yz/zx}$ and ${xy}$:
When $h$ is small, the hybridization between ${yz/zx}$ and ${xy}$ 
becomes strong and the ${yz/zx}$ bands have a gap around the Fermi level.
In the discussion for the ${yz/zx}$ band, hybridizations 
of the nearest $t_{yz/zx,3z^2-r^2}$ and $t_{yz/zx,x^2-y^2}$
are also important, especially for the 11 family.
In principles, these transfers must vanish from the mirror-plane symmetry 
of these orbitals, 
but, in fact, because of the symmetry braking due to the presence of 
the anion-$p$ component in the Wannier functions, 
the transfers exhibit finite values.
Now, when $h$ is large as in the case of FeTe, the symmetry 
tends to be largely broken so that $t_{yz/zx,3z^2-r^2}$ 
and $t_{yz/zx,x^2-y^2}$ become large. 
These transfers can induce a hybridization gap in the 
${yz/zx}$ bands, 
but a more important point is that those cause
a new gap formation at much lower energy than the Fermi level 
[around $-$1 eV, see Fig. \ref{fig:pdos}(c)]. 
As a result, this lower gap formation leads to closure of the pseudogap 
around the Fermi level.

Consequently, the 1111 family with smaller $h$ is closer to the band insulator 
due to the band splitting formed by the hybridization between the 3$d$ orbitals, 
regarded as a system with small carrier number at semimetallic electron and hole pockets.  
On the other hand, the 11 family has a large density of states over the Fermi level 
with fewer band splitting, which may make the electron correlation effect more efficient, 
because of larger effective carrier number, as in the formation of the Mott insulator in the middle of the band in the Hubbard model.   
We note that this remarkable trend from the semimetal to the ``half-filled" band emerges particularly 
for $yz/zx$ and $x^2-y^2$ orbitals in an {\it orbital selective} way.  
An {\it orbital selective crossover} from the band-insulating to Mott physics may specifically emerge 
in these $yz/zx$ and $x^2-y^2$ orbitals, while the other two orbitals keep more or less the band insulating character through this family variation. 
\\ 

The dispersion of the Fe-$3d$ band of the 11 family has larger three dimensionality.  
This larger dispersion in the 11 family is mainly due to the lack of the block layer including La and O, which makes the 
three-dimensional overlap of Fe-3$d$ orbital through the chalcogen-$p$ orbital larger.  
It appears to be triggered by the larger $h$ of the 11 family as well.  The larger $h$ means 
that the distance of the pnictogen or chalcogen to the other neighboring Fe layer in turn gets even smaller for the 11 family.  
Then this leads to a more efficient role of the chalcogen to bridge the two neighboring Fe layers to make 
larger dispersion for the 11 family. On the other hand, the distance between the pnictogen 
and the other neighboring Fe layer for the 1111 family is too far to make the appreciable three-dimensional dispersion.   \\ 
 
We now summarize the characteristic feature specific to the $d$ models with their family dependence.
The larger hybridization between Fe-3$d$ and $p$ orbitals of pnictogen for the 1111 compounds leads to 
larger extensions of the Wannier orbitals over the 11 family for the $d$ model, because the Wannier functions 
of the Fe-3$d$ orbitals more strongly mixes 
with the $p$ orbitals for the 1111 family. 
Still, the nearest-neighbor transfer for the 1111 family is not particularly large. Indeed it is 
even smaller than the other families including the 11 family, 
presumably because of large in-plane lattice constant.
The largest transfer ($\sim 0.4$ eV) appears between the two $xy$ orbitals 
because the orbitals are directed to the direction of the neighboring Fe sites, 
where the amplitude is 342, 315, 410, and 392 meV for LaFePO, LaFeAsO, FeSe, and FeTe, respectively.        

On the other hand, the largest next-nearest transfer shows little family dependence. 
Its value is 0.345 eV for LaFeAsO, 0.356 eV for BaFe$_2$As$_2$ and 
0.348 eV for FeSe and FeTe. 
These largest transfers are for the hopping between the diagonal elements between $yz$ or $zx$ orbitals.  
As a consequence of this trend of the nearest and next-nearest neighbor transfers, the 11 family 
is expected to show weaker geometrical 
frustration effects probed by the ratio of  
the next-nearest-neighbor to the nearest-neighbor transfers. 

Although the nesting picture predicts a similar periodicity of the antiferromagnetic order 
for all the families in contrast to the experimental variation, the contrast in the frustration parameter 
will offer a possibility of explaining the variation of the experimental periodicity in the  
magnetic order depending on the compounds 
from the viewpoint of the strong correlation physics.   
 
The effective Coulomb repulsion averaged over the Fe-3$d$ orbitals for the $d$ model is 
2.5, 2.8, 3.2, 4.2 and 3.4 for LaFeAsO, BaFe$_2$As$_2$, LiFeAs, FeSe and FeTe, respectively, 
while the overall bandwidth is around 4.5 eV in most cases except for LaFePO (5.1 eV) and LiFeAs (4.9 eV). 
This indicates that FeSe is located in much more strongly correlated region than the 1111 family, 
while the 122 and 111 families as well as FeTe occupy an intermediate region between the 1111 family and FeSe. 
In fact, FeSe has a substantially larger ratio of $\bar{U}/\bar{t}\sim 14$ as compared to 9 for LaFeAsO.  
An important origin of the larger 
$\bar{U}/\bar{t}$ for FeSe is the smaller extension of the Wannier orbitals ascribed to 
the smaller hybridization with the chalcogen-$p$ orbitals.  A smaller Wannier spread yields 
a larger bare Coulomb interaction itself.  Another origin of the difference is the poor screening 
in the case of the 11 family ascribed to the fewer screening channels in the absence of 
La-$f$ and O-$p$ bands. 
In fact, these bands are sources of efficient screening of the Coulomb interaction for the electrons on the iron-3$d$ orbitals in the 1111 family.  
The distance from the Fe layer to the pnictogen or chalcogen elements, namely the $h$ value itself contributes, 
where the larger distance makes the screening weaker for the 11 family. 

In other words, an overall trend of the family dependence of $U$ is explained in the following way: The bare Coulomb interaction $v$ is determined predominantly by the local environment and the extension of the Wannier orbital in the real space.  On the other hand, the screening effect measured from 
$\bar{U}/\bar{v}$ shown in Table \ref{tab:uv} is mainly determined from the band structure in the momentum space. In particular the presence of other bands close to the Fermi level is important, which efficiently screens the interaction of the Wannier orbitals of the target bands.  As a result, the 1111 family has weaker effective Coulomb interaction $\bar{U}$ as a combined effect of the efficient screening by many $p$ bands leading to smaller 
$\bar{U}/\bar{v}$, and the larger extension of the Wannier orbital leading to a smaller $\bar{v}$.   

Although the overall trend is understood from $h$, $h$ alone is not enough to explain all the details. 
There are some exceptions, and they are ascribed to other factors ({\it e.g.} number of screening channels, energy difference between 
occupied and unoccupied states). 
For example, FeSe has larger value of $U$ than FeTe, 
because the chalcogen-$p$ level is deeper (Sec.\ref{subsec:U}). 
The $U$ is not much different between LaFePO and LaFeAsO. 
This is because difference of the anion (P and As) radius partly cancels the difference in $h$. 
Larger $d$ bandwidth and higher La-$f$ level weaken screening in LaFePO, 
which also reduce the difference between the two compounds.

The $d$ model indicates that $U$ strongly depends on the orbital for the 1111 family 
while this anisotropy is relatively weaker for the 11 family.  In general, the ${x^2-y^2}$ orbital has 
the weakest $U$ because of the largest extension of the Wannier orbital extending to the pnictogen site.  
However, for FeTe, the orbital dependence of the extension of the Wannier orbitals is weak 
because of the weak covalency between Fe and Te.  Then, $U$ for the ${x^2-y^2}$ orbital ($\sim 3.6$ eV) 
is even larger than those for ${yz}$ and ${zx}$ orbitals $\sim 2.9$ eV in contrast to 
the prominently weak $U$ of the ${x^2-y^2}$ orbital for the 1111 family. \\

The larger magnetic ordered moment observed in the 122 and 11 families may basically be understood from the difference in $\bar{U}/\bar{t}$.  
The larger moment may also be ascribed to the lack of the pseudogap structure in FeTe, 
which makes the correlation effect larger, while the 1111 family is located in the region characterized by ``semimetal" with weaker correlation. 
The exchange interaction (Hund's rule coupling) $J$ averaged over orbitals is 0.39, 0.43, 0.51 and 0.47 eV for 
LaFeAsO, BaFe$_2$As$_2$, FeSe and FeTe, which has a trend of the orbital dependence similar to $U$. 

In the effective models containing chalcogen- or pnictogen-$p$ orbitals, namely $dp/dpp$ models, 
$U$ ranges from 4 eV for the 1111 family to 7 eV for FeSe. This family dependence partly comes from 
the family dependence of $v$, reflecting the family dependence of the extension of the Wannier orbitals as seen in Table \ref{tab:uv}. 
The orbital dependence is much weaker than in the case of the $d$ models and more or less isotropic correlation is justified. 
This indicates that the anisotropy of $U$ in the $d$ models arises from the individuality of the $p$ bands. 
Although the $dp/dpp$ models are quite complicated to analyze, they have several advantages. 
First, the isotropic correlation is convenient when we combine model calculations and LDA; 
in order to avoid double counting of the correlation effect considered in LDA, we have to introduce 
the so-called counter term to cancel the double counting in the model. In the $dp/dpp$ models, 
since the interaction parameters are almost isotropic (orbital independent), the counter term of the double counting 
is very simple, i.e., orbital independent.
Second, the $dp/dpp$ models may describe the antiferromagnetic state more appropriately especially for FeTe.
There, while the Fe-3$d$ bands are not so isolated, the exchange splitting is expected to be appreciable
since the magnetic moment is 2.2 $\mu_B$.
Another possible advantage of the $dp$ and $dpp$ model is, we can explicitly study the polarization effect of 
the anion-$p$ orbitals, which has been proposed to play an important role in the pairing mechanism.\cite{Sawatzky,Sawatzky2}

Now we have elucidated that this systematic variation of the parameters of the effective low-energy models from the 1111 to 11 families revealed in the present study is in many aspects originated and understood from the variation of $h$, resulting in the change from a strong covalency to a strong ionicity in the order from LaFePO, LaFeAsO, LiFeAs, BaFe$_2$As$_2$, FeSe to FeTe. The larger correlation parameter $\bar{U}/\bar{t}$ evolves basically in this order, while the frustration parameter $t'/t$ diminishes more or less in the same order and the three-dimensionality also evolves in this order, all of which are understood from this interplay of the covalency/ionicity.  The effect of covalency has already been discussed in the comparison of the band structure between LaFePO and LaFeAsO by Vildosola {\it et al.}.\cite{vildosola08} The present work has revealed that this key parameter $h$ controls in a more profound fashion the parameters of the effective low-energy models including the correlation effects and the geometrical frustration.  

Although the correlation effect was not clearly visible in the core level spectroscopy of LaFeAsO$_{1-x}$F$_x$,~\cite{Malaeb} this can be different for the 11 family in this respect of the difference in correlation amplitude. 
Recent soft X-ray photoemission measurement~\cite{Shin,Yamasaki} has elucidated the valence band spectra of FeSe$_{1-x}$ with $x\sim 0.08$. In particular, Yoshida {\it et al.} has observed, by the 140 eV photon source, a clear peak at the shoulder of the Fe-$3d$ band around 2 eV below the Fermi energy with a dip around 1 eV between the shoulder-like peak at $\sim 2$ eV and a rather sharp coherent structure around the Fermi level (see Fig.~1 of ref.~\citen{Shin}). Although the authors did not analyze in detail, this structure does not appear to be fully consistent with our density of states shown in Fig.~\ref{fig:dos}.  Instead, this structure primarily originated from Fe-$3d$ bands is reminiscent of the splitting of the coherent and incoherent part arising from the electron correlation effect.\cite{Suga}   
It would be intriguing to perform experiments for FeSe and FeTe to see whether the coherent-incoherent splitting as well as the satellite structure specific to the electron correlation effect exists or not in more detail. In addition, although FeSe has the largest 
$\bar{U}/\bar{t}$ 
in the series of the compounds, it is known to be paramagnetic metal with the superconductivity at low temperatures, while FeTe shows antiferromagnetic order with a large ordered moment, naively suggesting
a larger correlation effect in FeTe rather than in FeSe in contradiction with our obtained value of $U$. 
It might be related to the fact that it is difficult to synthesize the purely stoichiometric compound of FeSe.  It is important to control the stoichiometry experimentally for the purpose of clarifying the correlation effect expected in the 11 family.

In the core level spectroscopy of X-ray, it has been claimed that the onsite interaction $U$ is smaller than 2 eV,\cite{Shen_Xray} in apparent contradiction with the present model parameter values of the 1111 and 122 family, in the range of 2-3 eV.  We note that the ``$U$ value" speculated from the comparison between the photoemission data and the small cluster diagonalization without the screening by itinerant electrons should be taken with care. In fact, the random phase approximation taking into account 
the full polarization including iron-$3d$ electrons generates much smaller screened interaction 
$U\sim 0.9$ eV, roughly one third of the present cRPA estimates. 
Since the cluster diagonalization ignores the metallic screening, the ``$U$ value" consistent with the photoemission results should be similar to this full RPA value, while this fully screened ``$U$ value" turned out to be a substantial underestimate of the model parameter.  Therefore, the obtained model parameters and the overall photoemission data do not contradict each other, but are rather consistent. At the same time, it does not mean that the correlation effect is small, because the orbital fluctuations and orbital restructuring may be caused by the energy scale of the present $U$ obtained by the cRPA for the effective low-energy model. If the present interaction by the cRPA is comparable or larger than the bandwidth of the target bands, we expect substantial correlation effects. This is particularly true for the multi-orbital systems with screened interactions strongly dependent on the orbitals. In iron-based superconductors, this criterion for the substantially strong correlation appears to apply.   

The material dependence systematically revealed in this study will serve in clarifying the position of this series of compounds in the parameter space between the weakly correlated systems and the strongly correlated region governed by the Mott physics.  Although physical properties of LaFePO might be understood rather well as a weakly correlated system with the nesting scenario, the weak correlation picture becomes more and more problematic in the other compounds. In the 11 family, physical properties can be understood only by considering much wider energy range of the excitations away from the Fermi level in the order of the bandwidth, if one wishes to start from the perturbative picture. The local picture of the strong correlation becomes more applicable to 
the 11 family. It is remarkable that the ordered magnetic moment roughly increases when the effective Coulomb interaction increases. The mechanism of the superconductivity can also be pursued in this circumstance of the correlation effect, where the superconducting critical temperature $T_c$ appears to be optimized in this intermediate region, namely the region of a strong crossover or a possible quantum critical point.  An important novel aspect is that the orbital degeneracy and fluctuations of the five Fe-3$d$ orbitals may play a crucial role in physics of the iron-based superconductors in this intermediate region. 
 The $yz/zx$ and $x^2-y^2$ occupations are largely fluctuating in this respect and shows a strong crossover from the semimetals to the strongly correlated "half-filled" system with an orbital-selective crossover.   
 Since this region is hard to approach neither from the weak coupling nor strong coupling limits, these may be studied in the next step of the three-stage scheme, where we will continue to numerically solve the low-energy models by reliable low-energy solvers.  

There is an important caution when one performs model 
calculations by using the parameters determined in the present study:
The present parameters are derived for the real 
three-dimensional (3D) system and therefore the derived parameters 
 should be regarded as inputs for the 3D lattice-Fermion model.  
Conventionally, however, iron-based materials have been frequently studied 
as two-dimensional (2D) layer models.   
If one wishes to extract parameters for a purely 2D effective model, 
one has to introduce an additional treatment 
in the constrained calculation;   
when a layer being the target of the 2D model is specified and distinguished  
 from other layers, screening from the other layers  
should be included in the polarization calculations, with   
 excluding only the screening by the polarizations 
within the target layer itself.  
In the iron compounds, each layer is metallic, so  
this interlayer screening effect 
resulting from the  
treatment above could give nonnegligible corrections  
of the frequency-dependent dynamical screening 
to the presented parameters for the 3D system.  
Estimates of this correction are important  
in connecting the {\em ab initio} calculations with 
the 2D model analysis in a realistic way, which will be 
presented elsewhere in more detail. 

In the present formalism, the effective low-energy models of the $d$ or $dp/dpp$ types have been derived and proposed by eliminating the degrees of freedom originating from the other bands. In this procedure, we have employed LDA/GGA for obtaining the global band structure and the cRPA for calculating the screened interaction.  In principle, the low-energy electronic structure should be determined in a self-consistent way by considering the eliminated degrees of freedom again.  This possible feedback effect has been ignored in the present treatment.  This is justified under the circumstance that the bands of the eliminated degrees of freedom are well separated from the target bands near the Fermi level. The present iron superconductors appear to satisfy this condition rather well, while it is possible that the feedback gives a small but finite quantitative correction. Estimates of this correction is left for future studies.             
 
\acknowledgements    
We thank Yoshihide Yoshimoto,  
Taichi Kosugi, Takahiro Misawa, Yoshihiko Takano,
Antoine Georges, Silke Biermann, and Markus Aichhorn for useful discussions. 
We would like to thank financial support from 
the Next Generation Supercomputer Project, Nanoscience Program.  
A part of our computation has been done using 
the facilities of the Supercomputer Center, Institute for Solid 
State Physics, University of Tokyo. 
 
\vspace*{5mm}
{\it Note added in proof}\\
Very recently, N. Qureshi {\it et al.} (ArXiv:1002.4326) 
have reported that the magnetic moment of LaFeAsO is 0.63 $\mu_B$,  which is significantly larger than the values (0.36 $\mu_B$) previously reported in Ref. 10.  This is, however, still much smaller than the LSDA estimates.

\end{document}